\documentclass[11pt,tbtags,reqno]{amsart}

\usepackage{mathrsfs}
\usepackage{latexsym}
\usepackage{amsthm,amsopn,amsfonts,amssymb,epsfig,tabmac}
\numberwithin{equation}{section}

\makeatletter
\renewcommand{\subsubsection}{\@startsection
{subsubsection}
{3}
{0mm}
{\baselineskip}
{-0.5\baselineskip}
{\normalfont\normalsize\bfseries}}
\makeatother

\newtheorem{theorem}{Theorem}
\newtheorem{lemma}[theorem]{Lemma}
\newtheorem{proposition}[theorem]{Proposition}

\newtheorem{conjecture}[theorem]{Conjecture}

\newtheorem{corollary}[theorem]{Corollary}

\theoremstyle{remark}
\newtheorem{remark}[theorem]{Remark}
\newtheorem*{acknow}{Acknowledgments}

\def\la{{\lambda}}

\def\cal L{{\mathcal L}}

\def\Z{{\mathbb Z}}

\newcommand{\tcercle}[1]{\ensuremath{\setlength{\unitlength}{1ex}\begin{picture}(2.8,2.8)\put(1.4,1.4){\circle{2.8}\makebox(-5.6,0){#1}}\end{picture}}}
\newcommand{\gcercle}{\ensuremath{\setlength{\unitlength}{1ex}\begin{picture}(5,5)\put(2.5,2.5){\circle{5}}\end{picture}}}

\newcommand{\ds}{\displaystyle}
\newcommand{\mc}{\mathcal}

\newcommand{\versg}[1]{\ensuremath{\overleftarrow{\,\phantom{|}#1\,\phantom{|}}}}
\newcommand{\versd}[1]{\ensuremath{\overrightarrow{\,\phantom{|}#1\,\phantom{|}}}}
\newcommand{\aand}{\ensuremath{ \quad\textrm{and}\quad}}
\newcommand{\wwhere}{\ensuremath{ \quad\textrm{where}\quad}}
\let\d\partial
\let\n\noindent

\let\la\lambda
\let\La\Lambda
\let\Om\Omega

\let\ta\theta

\newcommand{\LL}{\ensuremath{\langle\!\langle}}
\newcommand{\RR}{\ensuremath{\rangle\!\rangle}}

\begin{document}

\title{Orthogonality of Jack polynomials in superspace}

\author{Patrick Desrosiers}
\address{Department of Mathematics and Statistics, The University of Melbourne, Parkville, Australia, 3010.}

\email{p.desrosiers@ms.unimelb.edu.au}

\address{New Address : Centre de recherches math\'ematiques,
Universit\'e de Montr\'eal,
P.O.\ Box 6128,
Centre-ville Station,
Montr\'eal (Qu\'ebec), Canada
H3C 3J7.
}
\email{desrosiers@crm.umontreal.ca}
\author{Luc Lapointe}
\address{Instituto de Matem\'atica y F\'{\i}sica, Universidad de
Talca, Casilla 747, Talca, Chile.}
\email{lapointe@inst-mat.utalca.cl }
\author{Pierre Mathieu}
\address{D\'epartement de physique, de g\'enie physique et
d'optique, Universit\'e Laval,  Qu\'ebec, Canada, G1K 7P4.}
\email{pmathieu@phy.ulaval.ca}

   \begin{abstract}{Jack polynomials in superspace,
orthogonal with respect to a ``combinatorial'' scalar product, are
constructed.  They are shown to coincide with the Jack polynomials
in superspace, orthogonal with respect to an ``analytical'' scalar product,
introduced in \cite{DLM3} as eigenfunctions of a supersymmetric quantum
mechanical many-body problem.  The results of this article rely on generalizing (to include an extra parameter)
the theory of classical symmetric
functions in superspace developed recently in \cite{DLM6}.}\end{abstract}


 \maketitle

\tableofcontents

\section{Introduction}

Jack polynomials, $J_\lambda(x;1/\beta)$, are symmetric functions of
commutative indeterminates $x=(x_1,\ldots,x_N)$ that generalize the
elementary ($\beta=\infty$), monomial ($\beta=0$), Schur
($\beta=1$), and zonal ($\beta=1/2$) symmetric functions. First
introduced in statistics by Jack \cite{Jack}, they were later
studied in algebraic combinatorics, in particular by Kadell \cite{Kadell},
Macdonald \cite{MacJack,Mac}, Stanley \cite{Stan}, and Knop and Sahi
\cite{Knop}.

The standard definition of the (monic) Jack polynomials is the
following \cite{Mac}: they are the unique functions such that
\begin{equation}\label{jackdef1}
(1)\quad
J_\lambda=m_\lambda+\sum_{\mu<\lambda}c_{\lambda\mu}(\beta)m_\mu
\qquad\text{and} \qquad (2)\quad
 \LL J_\lambda|J_\mu\RR_{\beta}\propto
\delta_{\lambda,\mu}\,,\end{equation} where $\lambda$ and $\mu$
stand for partitions of size not larger than $N$, $m_\lambda$ is the
monomial symmetric function, and $\mu<\lambda$ means that the
latter partition is larger than the former in the dominance
ordering.  The scalar product involved in (2) is of a combinatorial
nature.  On the basis of power-sum symmetric functions, it is defined
as
\begin{equation}\label{scalprodp}
\LL {p_\la} | {p_\mu }\RR_{\beta}:=\beta^{-\ell(\lambda)}z_\la
\delta_{\la,\mu}\, , \qquad {\rm where} \qquad
z_\la= \prod_i i^{m_i} m_i! \quad{\rm
if}\quad \la =(1^{m_1}2^{m_2}\cdots)\;. \end{equation}

However, alternative characterizations  of the Jack
polynomials exist. For instance, when the indeterminate $x_j$ is a complex
number lying on the unit circle and $\beta$ is a
nonnegative real number, one can introduce another scalar product
\cite{MacJack}:
\begin{equation}\label{scalprodphys}\langle
f(x)|g(x)\rangle_{\beta,N}=\prod_{1\leq j\leq N}\frac{1}{2\pi
\mathrm{i}} \oint \frac{ dx_j}{x_j}\prod_{\substack{1\leq k, l\leq
N\\k\neq l}}\left(1-\frac{x_k}{x_l}\right)^\beta \overline{f({
x})}\, g(x)\, ,\end{equation} where the bar denotes the complex
conjugation.    Then, it can be shown that Jack  polynomials are
the unique symmetric functions that satisfy
\begin{equation}\label{jackdef2}
(1)\quad
J_\lambda=m_\lambda+\sum_{\mu<\lambda}c_{\lambda\mu}(\beta)m_\mu
\qquad\text{and} \qquad (2')\quad \langle  J_\lambda
|J_\mu\rangle_{\beta,N}\propto
\delta_{\lambda,\mu}\,.\end{equation}
This  analytical scalar product is rooted in
the characterization of the Jack polynomials in terms of an eigenvalue problem; that is,
as the common
eigenfunctions of $N$ independent and commuting differential
operators that are self-adjoint with respect to  the scalar product
\eqref{scalprodphys}.  These operators are in fact the conserved
quantities of a well known $N$-body problem in quantum mechanics,
the trigonometric Calogero-Moser-Sutherland model.  Every orthogonal
and symmetric wave function of this model is proportional to a
particular  Jack polynomial \cite{Forrester,Lapointe}.

In this work, we provide an extension to the theory of classical
symmetric functions in superspace  \cite{DLM6} that  leads to a definition of Jack polynomials in superspace similar to (\ref{jackdef1}). By superspace, we
refer to a collection of variables $(x,\theta)=(x_1, \cdots x_N,
\ta_1, \cdots \ta_N)$, called respectively bosonic and fermionic (or
anticommuting or Grassmannian), and obeying the relations
\begin{equation}
x_ix_j=x_jx_i, \, \qquad
x_i\ta_j=\ta_jx_i\, \qquad {\rm and} \qquad \ta_i\ta_j=-\ta_j\ta_i \qquad (\Rightarrow \theta_i^2=0)\;.
\end{equation}
A function in superspace (or superfunction for short) is a function
of all these variables. It is said to be symmetric if it is
invariant  under the simultaneous interchange of $x_i
\leftrightarrow x_j$ and $\theta_i \leftrightarrow \theta_j$ for any
$i,j$.  Symmetric polynomials in superspace are naturally indexed by
superpartitions \cite{DLM1},
\begin{equation}\label{defsuperpart}
\Lambda:=(\La^a;\La^s)=(\Lambda_1,\ldots,\Lambda_m;\Lambda_{m+1},\ldots,\Lambda_{N})\,
,\end{equation} where $\La^a$ is a partition with distinct parts
(one of them possibly equal to zero), and $\La^s$ is an ordinary
partition.   Every symmetric polynomial in $x$ and $\theta$ can be
written as a linear combination of the following monomial functions
\cite{DLM1}:
\begin{equation} m_{\Lambda}=\sum_{\sigma\in S_{N}}\nolimits' \theta_{\sigma(1)} \cdots \theta_{\sigma(m)}x_1^{\Lambda_{\sigma(1)}} \cdots
x_m^{\Lambda_{\sigma(m)}} x_{m+1}^{\Lambda_{\sigma(m+1)}} \cdots
x_{N}^{\Lambda_{\sigma(N)}},
\end{equation} where the prime indicates that the  summation is
restricted to distinct terms. Power sums with $m$ fermions are given
by \cite{DLM6}
\begin{equation}
p_\La:=\tilde{p}_{\La_1}\ldots\tilde{p}_{\La_m}p_{\La_{m+1}}\cdots
p_{\La_N}\quad\text{with}\qquad
p_n:=m_{(n)}\quad\text{and}\quad\tilde{p}_k:=m_{(k;0)}\,
.\end{equation} In the article, we define
a simple
extension of the combinatorial scalar product \eqref{scalprodp}:
\begin{equation} \label{defscalprodcomb} \LL
{p_\La} | {p_\Om }\RR_\beta:=(-1)^{m(m-1)/2}z_\La
(\beta)\delta_{\La,\Om}\,,\quad
z_\La(\beta):=\beta^{-{\ell}(\La)}z_{\La^s}\, \end{equation} where
$\La$ is of the form \eqref{defsuperpart} and where $\ell(\La)$ is
the length of $\La$ (given by the length of $\La^s$ plus $m$).

Jack polynomials in superspace were presented  in \cite{DLM3} as the
orthogonal eigenfunctions  of a supersymmetric generalization of the
quantum mechanical $N$-body problem previously mentioned \cite{Brink,SS}.
In this case, the analytical scalar product reads \cite{DLM1}
\begin{equation}
\label{physca}\langle
A(x,\theta)|B(x,\theta)\rangle_{\beta,N}=\prod_{1\leq j\leq
N}\frac{1}{2\pi \mathrm{i}} \oint \frac{ dx_j}{x_j}\int
d\theta_j\,\theta_j\prod_{\substack{1\leq k, l\leq N\\k\neq
l}}\left(1-\frac{x_k}{x_l}\right)^\beta \overline{A({
x},{\theta})}\, B(x,\theta)\, ,
\end{equation} where the ``bar conjugation'' is defined
such that $ {\bar x}_j= 1/x_j$ and $\overline
{(\theta_{i_1}\cdots\theta_{i_m})}\theta_{i_1}\cdots\theta_{i_m}=1$.
Our main result here  is that these Jack polynomials in
superspace are also orthogonal with respect to the scalar product
(\ref{defscalprodcomb}); i.e., the two scalar products are compatible. The following theorem is an alternative
formulation of this statement.

\begin{theorem}
There exists a unique family of functions $\{ J_{\La}: \sum_i\La_i<N\}$
such that
 \begin{gather*}
(1)\quad J_{\Lambda} = m_{\Lambda} +\sum_{\Om < \La} c_{\La \Om
}(\beta) m_{\La}
\\
(2) \quad\LL J_{\La}| J_{\Om} \RR_{\beta} \propto
\delta_{\La,\Om} \quad \forall \, \La,\Om \qquad\mbox{or}\qquad (2')\quad\langle J_{\La}|
J_{\Om} \rangle_{\beta,N} \propto \delta_{\La,\Om} \quad \forall \, \La,\Om  \, ,
\end{gather*}
where  the ordering involved in  the triangular decomposition is the Bruhat
ordering on superpartition that will be defined in the next section.
\end{theorem}

 The article is organized as follows.  In Section~2, we summarize the theory of symmetric functions in
 superspace developed in \cite{DLM6}. We obtain a one-parameter deformation of the latter
 construction in Section~3.
 Section~4 is essentially  a review of relevant results concerning our previous
(analytical) construction  of  Jack polynomials in superspace.   It is
shown in Section 5 that these polynomials are also orthogonal with
respect to the product (\ref{defscalprodcomb}). Direct
non-trivial limiting cases (i.e., special values of the free
parameter or particular superpartitions) of this connection are
presented in Section~6.  This section also contains a discussion of
a duality transformation on the Jack superpolynomials, as well as a
conjectured expression for their normalization
constant.
    We present, as a concluding remark (Section 7), a precise conjecture concerning
 the existence of Macdonald polynomials in  superspace.

In this work, we have relied heavily on the seminal paper
\cite{Stan}, and  on Section~VI.10 of \cite{Mac}, without always
giving these references complete credit in the bulk of the paper.

\begin{remark}The terms ``superanalogs of Jack polynomials'',
``super-Jack polynomials'' and ``Jack superpolynomials'' have also been
used in the literature for somewhat different polynomials. In
\cite{Serg}, superanalogs of Jack polynomials designated the
eigenfunctions of the CMS Hamiltonian constructed from the root
system of the Lie superalgebra $su(m,N-m)$ (recall that to any root
system corresponds a CMS model \cite{OP}). The same objects are
called super-Jack polynomials in \cite{VS}.
   But we
stress that such a Hamiltonian does not contain anticommuting
variables, so that the resulting eigenfunctions are quite different
from our Jack superpolynomials.  Notice also that in
\cite{DLM1,DLM2}, we used the term ``Jack superpolynomials'' for
eigenfunctions of the supersymmetric extension of  the trigonometric
Calogero-Moser-Sutherland model that decompose triangularly in
the monomial basis. However, these are not orthogonal. The
construction of orthogonal Jack superpolynomials was presented in
\cite{DLM3} and from now on, when we refer to ``Jack
superpolynomials'', or equivalently, ``Jack polynomials in
superspace'', we refer to the orthogonal ones.\end{remark}

\section{Notation and background}
$\La\vdash(n|m)$ indicates that the
superpartition $\Lambda=(\Lambda_1,\dots,\Lambda_m;\Lambda_{m+1},
\dots,\Lambda_N)$ is of bosonic
degree $n=|\La|=\Lambda_1+\cdots+\Lambda_N$ and of fermionic degree
$m=\overline{\underline{\La}}$  respectively (oberve that the bosonic
and fermionic degree refer to the respective
degrees in $x$ and $\theta$ of
$m_{\Lambda}$). To every
superpartition $\La$, we can associate a unique partition $\La^*$
obtained by deleting the semicolon and reordering the parts in
non-increasing order.  A superpartition $\La=(\La^a;\La^s)$ can be
viewed as the partition $\Lambda^{*}$ in which every part of $\La^a$
is circled.
 If a part ${\La^a}_j=b$ is equal to at least one part of
$\La^s$, then we circle the leftmost $b$ appearing in $\La^*$. We
shall use $C[\Lambda]$ to denote this special notation.

To each $\La$, we associate the diagram, denoted by $D[\La]$,
obtained by first drawing the Ferrers' diagram associated to
$C[\La]$, that is, by drawing  a diagram with ${C[\La]}_1$ boxes in
the first row, ${C[\Lambda]}_2$ boxes in the second row and so
forth, all rows being left justified.  If, in addition, the integer
${C[\La]}_j=b$ is circled, then we add a circle at the end of the
$b$  boxes in the $j$-th row.

This representation offers a very
natural way to define a
conjugation operation. The  conjugate of a
superpartition $\La$, denoted by $\La'$, is obtained by
interchanging the rows and the columns in the diagram $D[\La]$. We
can thus write $D[\La']=(D[\La])^{\mathrm{t}}$ where $\mathrm{t}$
stands for the transposition operation. For instance, we have
\begin{equation} \label{exdia}D([3,0;4,3] )={\tableau[scY]{&&&\\&&&\bl\tcercle{}\\&&\\\tcercle{}\bl\\ }}\,  \qquad \implies \qquad
D([3,0;4,3])^t={\tableau[scY]{&&&\bl\tcercle{}\\&&\\&&\\&\bl\tcercle{}\\
}}\,
\end{equation}
meaning that $(3,0;4,3)'= (3,1;3,3)$.

We  now formulate the Bruhat ordering on superpartitions. Recall that two partitions
$\lambda$ and $\mu$ of $n$ are such that $\lambda$ dominates $\mu$
iff $\lambda_1+ \cdots+ \lambda_i \geq \mu_1+ \cdots+ \mu_i$ for all
$i$. The Bruhat ordering on superpartitions of $(n|m)$ can then be described
most simply as: $\Lambda \geq \Omega$ iff $\Lambda^* > \Omega^*$ or
$\Lambda^* = \Omega^*$ and ${\rm sh}(D[\Lambda]) \geq {\rm
sh}(D[\Omega])$, where ${\rm sh}(D[\Lambda])$ is the shape
(including circles) of the diagram $D[\Lambda]$ (see \cite{DLM6}
for the connection between this ordering and the usual Bruhat
ordering on superpartitions). With this
definition, it is then obvious that $\Lambda \geq \Omega$ iff
$\Omega' \geq \Lambda'$.

We denote by $\mathscr{P}^{S_\infty}$
the ring of symmetric functions in superspace with coefficients in $\mathbb Q$.
A basis for its subspace of homogeneous degree $(n|m)$ is given
by $\{ m_{\Lambda}\}_{\Lambda \vdash (n|m)}$ (now considered to be functions
of an infinite number of variables).
In this ring, the elementary $e_n$, homogeneous $h_n$, and power sum $p_n$
symmetric functions possess fermionic counterparts which are
obtained trough the following  generating functions:
\begin{align}\label{defEHP}
E(t,\tau)&:=\sum_{n=0}^\infty t^n(e_n +\tau\tilde{e}_n)
=\prod_{i=1}^\infty (1+t x_i + \tau \theta_i)\,
\\
 H(t,\tau)&:=\sum_{n=0}^\infty
t^n(h_n +\tau\tilde{h}_n )=\prod_{i=1}^\infty \frac{1}{1-t x_i -
\tau \theta_i}\,
\\
P(t,\tau)&:=\sum_{n\geq 1}(t^n\,p_n +\tau n t^{n-1}
\,\tilde{p}_{n-1}) =\sum_{i=1}^\infty \frac{t x_i + \tau
\theta_i}{1-t x_i - \tau \theta_i}\, ,
\end{align}
 where $\tau$ is an anticommuting parameter ($\tau^2=0$).  To be more explicit, this leads to
\begin{equation}
\tilde e_n = m_{(0;1^n)}\;, \qquad \tilde h_n = \sum_{\Lambda \vdash
(n|1)} (\Lambda_1+1)\, m_{\Lambda}\;, \qquad \tilde p_n =
m_{(n;0)}\;.
\end{equation}
This construction furnishes three multiplicative  bases $f_\La$ of
$\mathscr{P}^{S_\infty}$,
\begin{equation}
f_\La:=\tilde{f}_{\La_{1}} \cdots \tilde{f}_{\La_{m}}
{f}_{\La_{m+1}}\cdots {f}_{\La_{N}} \, ,\end{equation} where $f$ is
either $e,h$ or $p$.

With $(y_1,y_2,\dots,\phi_1,\phi_2,\dots)$ representing another set
of bosonic and fermionic variables (with the additional
understanding that $\phi_i \theta_j=-\theta_j \phi_i$), the
generalized Cauchy formula is shown to satisfy
\begin{equation}\label{defK} \prod_{
i,j}{(1-x_iy_j-\theta_i\phi_j)}^{-1}  = \sum_{\La}
\versg{m_\La(x,\theta)}\versd{h_\La(y,\phi)}=\sum_{\La}z_\La^{-1}
\versg{p_\La(x,\theta)}\versd{p_\La(y,\phi)} \;,\end{equation} where
\begin{equation} z_\La:=z_{\La^s}=  \prod_i i^{m_i} m_i! \qquad{\rm if}\qquad \La^s=(1^{m_1}2^{m_2}\cdots)\;,
\end{equation}
The arrows are used to encode signs resulting from reordering the
fermionic variables: if the fermionic degree of a polynomial $f$ in
superspace is $m$, then $\versg{f}=(-1)^{m(m-1)/2}f$ and
$\versd{f}=f$.

\section{One-parameter deformation of the
scalar product and the homogeneous basis}

Let $\mathscr{P}^{S_\infty}(\beta)$ denote the ring of symmetric
functions in superspace with coefficients in $\mathbb{Q}(\beta)$,
i.e., rational functions in $\beta$. We first introduce the mapping,
\begin{equation}\LL\,\cdot \, |\, \cdot\,\RR_{\beta} \,:
\,\mathscr{P}^{S_\infty}(\beta)\times
\mathscr{P}^{S_\infty}(\beta)\longrightarrow\mathbb{Q}(\beta)\end{equation}
defined by (\ref{defscalprodcomb}).  This bilinear
form can easily be shown to be a scalar product (using an argument
similar to the one given in \cite{DLM6} in the case
$\beta=1$).

We next introduce an endomorphism  that generalizes the involution
$\hat{\omega}$ of \cite{DLM6}, and which extends a known endomorphism
in symmetric function theory. It is defined on the power sums
as:
\begin{equation}\label{defendobeta}\hat{\omega}_\alpha(p_n)=(-1)^{n-1}\,\alpha\,
p_n\quad\mbox{and}\quad
\hat{\omega}_\alpha(\tilde{p}_{n})=(-1)^{n}\,\alpha\,\tilde{p}_{n}
\, , \end{equation} where $\alpha$ is some unspecified parameter.
This implies
\begin{equation}  \label{involugen}
\hat{\omega}_\alpha(p_\Lambda)=\omega_\La(\alpha)\,p_\Lambda
\quad\mbox{with} \quad \omega_\La(\alpha):= \alpha^{\ell(\Lambda)}
(-1)^{|\La|-\overline{\underline{\La}}+{\ell}(\La)}\,.\end{equation}
Notice that $\hat{\omega}_1\equiv \hat{\omega}$.  This
homomorphism is still self-adjoint, but it is now
neither an involution
($\hat{\omega}_\alpha^{-1}=\hat{\omega}_{\alpha^{-1}}$) nor an
isometry ($\|\hat{\omega}_\alpha p_\La \|^2=z_\La(\beta/\alpha^2)$).
Note also that
\begin{equation}\label{rels}
z_\La(\beta)\omega_\La(\beta)=z_\La\omega_\La\aand
z_\Lambda (\beta)^{-1} \omega_\Lambda(\beta^{-1}) = z_\La^{-1}
\omega_\Lambda\, .\end{equation}

We now extend the Cauchy kernel introduced in \eqref{defK}.

\begin{theorem} \label{cauchyppbeta} One has
 \begin{equation}
K^\beta(x,\ta;y,\phi):=\prod_{i,j}\frac{1}{(1-x_iy_j-\theta_i\phi_j)^{\beta}}\,
=\sum_{\La}z_\La(\beta)^{-1}
\versg{p_\La(x,\theta)}\versd{p_\La(y,\phi)}\,.
\end{equation}
\end{theorem}
\begin{proof}
Starting from
\begin{equation}\prod_{i,j}\frac{1}{(1-x_iy_j-\theta_i\phi_j)^{\beta}}=\exp\Big\{\beta\sum_{i,j}
\ln\Big[(1-x_iy_j-\theta_i\phi_j)^{-1}\Big]\Big\} \, ,\end{equation}
the above identity can be obtained straightforwardly by  proceeding as
in the proof of Theorem~33 of \cite{DLM6}. \end{proof}

\begin{remark} The inverse of $K^\beta$ satisfies:
\begin{equation}
 K(-x,-\ta;y,\phi)^{-\beta}=\prod_{i,j}(1+x_iy_j+\theta_i\phi_j)^{\beta}\,
=\sum_{\La} z_\La(\beta)^{-1}\omega_\La
\versg{p_\La(x,\theta)}\versd{p_\La(y,\phi)} \; ,\end{equation}
which is obtained by using
\begin{equation}p_\La(-x,-\theta)=(-1)^{|\La|+\overline{\underline{\La}}}\,
p_\La(x,\theta)\aand
  z_\La(-\beta)=(-1)^{\ell(\La)}z_\La(\beta)\,.\end{equation}
Notice also
the simple relation between the  kernel $K$ of \cite{DLM6} (equal to $K^\beta$ at $\beta=1$) and its $\beta$-deformation
\begin{equation}
K^{\beta}(x,\ta;y,\phi)=\hat{\omega}_\beta K(-x,-\ta;y,\phi)^{-1}
\, ,\end{equation} where it is
understood that $\hat{\omega}_\beta$ acts either on $(x,\theta)$ or
on $(y,\phi)$.
\end{remark}

\begin{corollary}
\label{Kernelbeta}  $K^\beta(x,\theta;y,\phi)$ is a reproducing
kernel in the space of symmetric superfunctions with rational
coefficients in $\beta$:
\begin{equation}\LL\,K^\beta(x,\theta;y,\phi)\,|\,f(x,\ta)\,\RR_\beta=f(y,\phi)\,
,\quad\mbox{for all}\quad f\in\mathscr{P}^{S_{\infty}}(\beta)\,
.\end{equation}\end{corollary}

Paralleling the construction of the function $g_n$ in Sect.~VI.10 of
\cite{Mac}, we now introduce a $\beta$-deformation of the bosonic
and fermionic complete homogeneous symmetric functions, respectively
denoted as $g_n(x)$ and $\tilde{g}_n(x,\theta)$ (the
$\beta$-dependence being implicit). Their generating function is
\begin{equation}\label{generatriceg} G(t,\tau;\beta):=\sum_{n\geq
0}t^n[g_n(x)+\tau \tilde{g}_n(x,\theta)]=\prod_{i\geq
1}\frac{1}{(1-tx_i-\tau\theta_i)^\beta}\, .\end{equation} Clearly,
$g_n=h_n$ and $\tilde{g}_n=\tilde{h}_n$ when $\beta=1$.  As usual,
we define
\begin{equation}\label{defgla}
g_\La:=\tilde{g}_{\La_{1}} \cdots \tilde{g}_{\La_{m}}
{g}_{\La_{m+1}}\cdots g_{\La_{N}} \, . \end{equation}

\begin{proposition}\label{propkernelmg}One has  $\ds
K^\beta(x,\theta;y,\phi)=\sum_{\La} \versg{m_\La(x,\theta)}
\versd{g_\La(y,\phi)}$.
\end{proposition}
\begin{proof}The proof is similar to that of Proposition 38 of \cite{DLM6}.\end{proof}

\begin{corollary}One has
\begin{equation}\label{genp} g_n=\sum_{\La\vdash(n|0)}z_\La(\beta)^{-1} p_\La
\aand \tilde{g}_n=\sum_{\La\vdash(n|1)}z_\La(\beta)^{-1} p_\La\,
.\end{equation}
\end{corollary}
\begin{proof}On the one hand,
\begin{equation}G(t,0;\beta)=\sum_{n\geq 0}t^n
g_n(x)=K^\beta(x,0;y,0)\Big|_{y=(t,0,0,\ldots)}\;.\end{equation}
The previous proposition and Theorem~\ref{cauchyppbeta} imply
\begin{equation}\sum_{n\geq0}t^n g_n=\sum_{\la}t^{|\la|}z_\la(\beta)^{-1}p_\la\quad\Longrightarrow\quad
g_n=\sum_{\la\vdash n}z_\la(\beta)^{-1}p_\la\, .\end{equation} On
the other hand, \begin{equation}\partial_\tau
G(t,\tau;\beta)=\sum_{n\geq 0}t^n
\tilde{g}_n(x,\theta)=K^\beta_+(x,\theta;y,\phi)\Big|_{\substack{y=(t,0,0,\ldots)\\
\phi=(-\tau,0,0,\ldots)}}\;. \end{equation}
 Hence
\begin{equation}\sum_{n\geq 0} t^n \tilde{g}_n=\sum_{\La,\overline{\underline{\La}}=1}t^{|\La|}z_\La(\beta)^{-1}p_\La\quad\Longrightarrow\quad
\tilde{g}_n=\sum_{\La\vdash(n|1)}z_\La(\beta)^{-1}p_\La\,
,\end{equation} as claimed.
\end{proof}

Applying
$\omega_{\beta^{-1}}$ on equation (\ref{genp}), simplifying with the help of
\eqref{rels}, and then
using \cite{DLM6}
\begin{equation}
e_n = \sum_{\Lambda \vdash (n|0)} z_{\Lambda}^{-1} \omega_{\Lambda} p_{\Lambda}
\qquad {\rm and} \qquad
\tilde e_n = \sum_{\Lambda \vdash (n|1)} z_{\Lambda}^{-1} \omega_{\Lambda} p_{\Lambda} \, ,
\end{equation}
we get \begin{equation}\label{dualityge}
\hat{\omega}_{\beta^{-1}} (g_n)=e_n\aand  \hat{\omega}_{\beta^{-1}}
(\tilde{g}_n)=\tilde{e}_n\, .\end{equation}
Or equivalently,
\begin{equation}
g_n = \hat{\omega}_{\beta} (e_n) \aand  \tilde g_n = \hat{\omega}_{\beta}
(\tilde{e}_n)\, .\end{equation}

\begin{lemma}\label{cauchybasesbeta}Let $\{u_\La\}$ and $\{v_\La\}$ be two
bases of $\mathscr{P}^{S_{\infty}}$.  Then
\begin{equation} \label{eqKN}
K^\beta(x,\theta;y,\phi)=\sum_\La
\versg{u_\La(x,\theta)}\,\versd{v_\La(y,\phi)}\quad\Longleftrightarrow\quad\LL
\versg{u_\La}|\versd{v_\La}\RR_\beta=\delta_{\La,\Om}\,
.\end{equation} \end{lemma}
\begin{proof}
The proof is identical to the one when the $\ta_i$ variables are not present
(see \cite{Mac} (I.4.6)).\end{proof}

This
immediately implies the following.
\begin{corollary}\label{corodualitygm}
The set $\{g_\La\}_\La$ constitutes a basis of
$\mathscr{P}^{S_\infty}(\beta)$ dual to that of the monomial
basis in superspace;
that is,
\begin{equation} \LL \versg{\,g_\La} | \versd{m_\Om
}\RR_\beta=\delta_{\La,\Om}\, .\end{equation}\end{corollary}

We will need in the next section to make explicit the distinction
between an infinite and a finite number of variables.   Therefore, we
also let
\begin{equation}\LL\,\cdot \, |\, \cdot\,\RR_{\beta,N} \,:
\,\mathscr{P}^{S_N}(\beta)\times
\mathscr{P}^{S_N}(\beta)\longrightarrow\mathbb{Q}(\beta)\end{equation}
where $\mathscr{P}^{S_N}$ is the restriction of  $\mathscr{P}^{S_\infty}$ to $N$ variables, defined by requiring that the bases
$\{ g_\La\}_{\ell(\La)\leq N}$ and $\{ m_\La\}_{\ell(\La)\leq N}$ be dual to each other:
\begin{equation} \label{defscalprodcombN} \LL
\versg{g_\La} | \versd{m_\Om }\RR_{\beta,N}:=\delta_{\La,\Om}
\, ,\end{equation}
whenever $\ell(\La)$ and $\ell(\Om)$ are not larger than $N$.
From this definition, it is thus obvious that
\begin{equation} \label{degN}
\LL\,f_1^{(N)} \, |\, f_2^{(N)} \,\RR_{\beta,N}=\LL\,f_1 \, |\,
f_2\,\RR_{\beta}
\end{equation}
if $f_1$ and $f_2$ are elements of the ring of
symmetric functions in superspace
of bosonic degrees
smaller than $N$, and if $f_1^{(N)}$ and $f_2^{(N)}$ are their
respective restriction to $N$
variables.  This is because $f_1$ and $f_1^{(N)}$ (resp. $f_2$ and
$f_2^{(N)}$ ) then
have the same expansion in terms of the $g$ and $m$ bases.  Note that
with this definition, we have that
\begin{equation}
K^{\beta,N} = \sum_{\ell(\La) \leq N} g_{\La}(x,\theta)\, m_{\La}(y,\phi) \, ,
\end{equation}
where $K^{\beta,N}$ is the restriction of $K^{\beta}$ to $N$ variables and
where $(x,\theta)$ and $(y,\phi)$ stand respectively for
$(x_1,\dots,x_N,\theta_1,\dots,\theta_N)$
and $(y_1,\dots,y_N,\phi_1,\dots,\phi_N)$.

We complete this section by displaying a relationship
between the $g$-basis elements and the bases of monomials and homogeneous polynomials.

\begin{proposition}\label{propgn}Let
$n_\La!:=n_{\La^s}(1)!\,n_{\La^s}(2)!\cdots$, where $n_{\La^s}(i)$ is the multiplicity of $i$ in $\La^s$,  and
\begin{equation}\binom{\beta}{n}:= \frac{(\beta)_n }{n!}
\;, \qquad (\beta)_n :=  \beta(\beta-1)\cdots(\beta-n+1)
.\end{equation}

Then
\begin{eqnarray}
\label{jnenm}g_n&=& \sum_{\La\vdash (n|0)}\prod_i \binom{
   \beta+\La_i-1}{
   \La_i }
m_\La\, =  \sum_{\La\vdash (n|0)}
\frac{(\beta)_{\ell(\La)}}{n_\La !} \,h_\La\, ,\\
\label{jnoenm}\tilde{g}_n&=& \sum_{\La\vdash
(n|1)}(\beta+\La_1)\prod_i \binom{
   \beta+\La_i-1}{
   \La_i }
m_\La\, = \sum_{\La\vdash
(n|1)}\frac{(\beta)_{\ell(\La)}}{n_\La!}\,h_\La\, .
 \end{eqnarray}
\end{proposition}
\begin{proof}We start with  the generating
function (\ref{generatriceg}). The product on the right hand side
can also be written as
\begin{eqnarray}
\lefteqn{\prod_{i\geq 1} \sum_{k\geq 0}(-1)^k\binom{-\beta }{k}
  (tx_i+\tau\theta_i)^k =}\cr
&&\prod_{i\geq 1}\left[ \sum_{k\geq 0}\binom{
   \beta+k-1}{k}
(tx_i)^k+\tau \theta_i \sum_{k\geq 1}k\binom{
   \beta+k-1}{
   k }(tx_i)^{k-1}\right]
\end{eqnarray}
After some easy manipulations, (\ref{generatriceg}) then becomes
\begin{equation}
G(t,\tau; \beta) = \sum_{n\geq 0}t^n\left[\sum_{\la\vdash n}
\prod_i\binom{
   \beta+\la_i-1}{\la_i }m_\la\,+\tau \sum_{\La\vdash (n|1)}(\beta+\La_1)\prod_i\binom{ \beta+\La_i-1}{   \La_i}\,m_\La\right]\end{equation} and  the first equality in the two formulas (\ref{jnenm}) and (\ref{jnoenm}) are
seen to hold.

To prove the remaining two formulas, we use the generating function
of the homogeneous symmetric functions and proceed as follows:
\begin{eqnarray}
\prod_i(1-tx_i-\tau\theta_i)^{-\beta} &=& \left( 1+\sum_{m\geq 1}t^m h_m+\tau \sum_{n\geq 0}
t^n\tilde{h}_n
\right)^\beta\cr
&=& \sum_{k\geq 0} \left(\begin{array}{c}
   \beta \\
   k \\
\end{array}\right)
\left( \sum_{m\geq 1}t^m h_m+\tau \sum_{n\geq 0} t^n\tilde{h}_n
\right)^k\cr &=&\sum_{n\geq 0}\sum_{\la\vdash
n}t^n\frac{(\beta)_{\ell(\la)}}{n_\la !}\, h_\la +\tau \sum_{m\geq
0}\sum_{\la\vdash m}t^m\frac{(\beta)_{\ell(\la)+1}}{\la !}\,
h_\la\sum_{n\geq 0}t^n \tilde{h}_n\cr &=&\sum_{n\geq
0}t^n\left[\sum_{\La\vdash (n|0)}\frac{(\beta)_{\ell(\La)}}{n_\La
!}\, h_\La +\tau \sum_{\La\vdash
(n|1)}\frac{(\beta)_{\ell(\La)}}{n_{\La^s} !}\, h_\La\right]
\end{eqnarray}from which the desired expressions can be obtained.
\end{proof}


\section{Jack polynomials in superspace: analytical characterization}

We review the main properties of Jack superpolynomials as they were defined in \cite{DLM3}.
 The section is completed with the presentation of a technical lemma to be used in Section 6.  All the results of this
section are independent of those of Section 3.

First, we define a scalar product in $\mathscr{P}$, the ring of
polynomials in superspace in $N$ variables. Given
\begin{equation}
\Delta(x) = \prod_{1\leq j<k \leq N} \left[ \frac{x_j-x_k}{x_j x_k}
\right] \, ,
\end{equation}
$\langle\cdot|\cdot\rangle_{\beta,N}$ is defined (for $\beta$ a
positive integer) on the basis elements of $\mathscr{P}$ as
\begin{equation} \label{scalarp}
\langle \, \theta_I x^{\lambda}|\theta_{J}x^{\mu} \,\rangle_{\beta,N}
= \left\{
\begin{array}{ll}
{\rm C.T.} \left[\Delta^{\beta}(\bar x) \Delta^{\beta}(x)
\bar{x}^{\mu}
x^{\lambda}\right] & {\rm if~} I=J\, , \\
0 & {\rm otherwise}\, .
\end{array}  \right.
\end{equation}
where  ${\bar x}_i= 1/x_i$, and  where ${\rm C.T.}[E]$ stands for
the constant term of the expression $E$. (This is another form of the
scalar product
(\ref{physca}). More precisely, the latter is the analytic deformation of the former for all values of $\beta$.) This gives our first characterization of the Jack
superpolynomials.
\begin{proposition}\cite{DLM3}
\label{charac1}
   There exists a unique basis $\{J_\Lambda\}_\La$ of $\mathscr{P}^{S_N}$
such that \begin{equation*} (1)\quad J_{\Lambda} = m_{\Lambda}
+\sum_{\Om < \La} c_{\La \Om }(\beta) m_{\La}\qquad\text{and}\qquad
(2')\quad \langle J_{\La}| J_{\Om}\rangle_{\beta,N} \propto
\delta_{\La,\Om}.\end{equation*}
\end{proposition}

In order to present the other characterizations, we need to introduce
the Dunkl-Cherednik operators (see \cite{Baker} for instance):
\begin{equation}\label{Dunk}{\mc{D}}_j:=x_j\partial_{x_j}+\beta\sum_{k<j}{\mc{O}}_{jk}+\beta\sum_{k>j}{\mc{O}}_{jk}-\beta(j-1)\,
,\end{equation} where
\begin{equation}{\mc{O}}_{jk}=\left\{\begin{array}{ll}
\ds\frac{x_j}{x_j-x_k}(1-{K}_{jk})\,,&k<j\, ,\\
\ds\frac{x_k}{x_j-x_k}(1-{K}_{jk})\,,&k>j\, .
\end{array}\right.\end{equation}
Here $K_{jk}$ is the operator that exchanges the variables
$x_j$ and $x_k$:
\begin{equation}
K_{jk}f(x_j,x_k,\theta_j,\theta_k)=f(x_k,x_j,\theta_j,\theta_k)\, .
\end{equation} The Dunkl-Cherednik operators can be used to define two families
of operators that preserve the elements of homogeneous
degree $(n|m)$ of $\mathscr P^{S_N}$:
\begin{equation}\label{defHI} \mathcal{H}_r:=\sum_{j=1}^N {{\mc{D}}_j^{\,
r}}\,\aand \mathcal{I}_s:=\frac{1}{(N-1)!}\sum_{\sigma\in
S_N}\mc{K}_\sigma \Bigl( \theta_1\partial_{\theta_1}{\mc{D}}_1^{\,
s} \Bigr)\mc{K}_{\sigma}^{-1} \, ,\end{equation} for $r\in
\{1,2,3,\dots,N \}$ and $s\in \{0,1,2,\dots,N-1\}$ and where $\mathcal K_{\sigma}$ is built out of the operators
  $\mathcal K_{jk}$ that exchange $x_j \leftrightarrow x_k$
and  $\theta_j \leftrightarrow \theta_k$ simultaneously:
\begin{equation}
\mc{K}_{i,i+1}:=\kappa_{i,i+1}K_{i,i+1}\qquad {\rm where }\qquad
\kappa_{ij}f(x_i,x_j,\theta_i,\theta_j)=f(x_i,x_j,\theta_j,\theta_i)\, \;.\end{equation}

The operators $\mathcal{H}_r$ and $\mathcal{I}_s$ are mutually commuting when restricted to $\mathscr{P}^{S_N}$; that is, \begin{equation}[ \mathcal{H}_r,
\mathcal{H}_s]f= [ \mathcal{H}_r, \mathcal{I}_s]f= [ \mathcal{I}_r,
\mathcal{I}_s]f=0
  \qquad \forall \, r,s \, ,
\end{equation} where $f$ represents an arbitrary polynomial in
$\mathscr{P}^{S_N}$. Since they are also symmetric with respect to
the scalar product $\langle\cdot|\cdot\rangle_{\beta}$ and have,
when considered as a whole, a non-degenerate spectrum, they provide
our second characterization of the Jack superpolynomials.
\begin{proposition} \cite{DLM3}
\label{charac2}
  The Jack superpolynomials $\{J_{\La}\}_{\La}$
are the unique common eigenfunctions of the $2N$ operators
$\mathcal{H}_r$ and $\mathcal{I}_s$, for $r\in \{1,2,3,\dots,N \}$
and $s\in \{0,1,2,\dots,N-1\}$.
\end{proposition}

We will now define two operators that play a special role in our study:
\begin{equation}\label{defHIs}
\mathcal{H}:=\mathcal{H}_2+ \beta (N-1)\mathcal{H}_1-\mathrm{cst}
\aand \mathcal{I}:=\mathcal{I}_1\, ,\end{equation} where
$\mathrm{cst}=\beta N(1-3N-2N^2)/6$.  When acting on symmetric
polynomials in superspace, the explicit form of $\mc{H}$ is simply
\begin{equation}
\mc{H}=\sum_i (x_i \partial_{x_i})^2+\beta
\sum_{i<j}\frac{x_i+x_j}{x_{i}-x_j}(x_i
\partial_{x_i}-x_j\partial_{x_j})-2\beta
\sum_{i<j}\frac{x_ix_j}{(x_i-x_j)^2}(1-\kappa_{ij})\, .
\end{equation}

The operator $\mc{H}$ is the Hamiltonian of the supersymmetric form of  the
trigonometric Calogero-Moser-Sutherland model (see
Section 1); it can be written in terms of two fermionic operators
$\mathcal{Q}$ and $\mathcal{Q}^\dagger$ as
\begin{equation}
\mathcal{H}= \mathcal{Q} \mathcal{Q}^\dagger+ \mathcal{Q}^\dagger\mathcal{Q} \, ,\end{equation}
where
\begin{equation} \mathcal{Q}:=\sum_i\theta_ix_i\partial_{x_i} \aand
\mc{Q}^\dagger=\sum_i\partial_{\theta_i}\left(x_i\partial_{x_i}
+\beta \sum_{j\neq i}\frac{x_i+x_j}{x_i-x_j}\right)\,
,\end{equation} so that $\mc{Q}^2= ( \mc{Q}^\dagger)^2=0$.
Physically, $\mc{Q}$ is seen as creating fermions while
$\mc{Q}^\dagger$ annihilates them.
 A state (superfunction) which is annihilated by the fermionic
operators is called supersymmetric.  In the case of
polynomials in superspace, the only supersymmetric state is the identity.

\begin{remark} The  Hamiltonian $\mc{H}$
has  an elegant differential geometric interpretation  as a
Laplace-Beltrami operator.  To understand this assertion,
consider  first the real Euclidian space $\mathbb{T}^N$, where
$\mathbb{T}=[0,2\pi)$.  Then, set $x_j=e^{\mathrm{i}t_j}$ for
$t_j\in\mathbb{T}$, and
 identify the Grassmannian  variable
$\theta_i$ with the differential form $dt_i$.  This allows us to
rewrite the scalar product \eqref{physca} as a Hodge-de~Rham
product involving complex differential forms; that is,
\begin{equation} \langle
A(t,\theta)|B(t,\theta)\rangle_{\beta,N}\sim
\int_{\mathbb{T}^N}\overline{A(t,dt)}\wedge \ast B(t,dt)\,
,\end{equation}where the bar denotes de complex conjugation and
where  the Hodge duality operator $\ast$ is formally defined by
\begin{equation}
A(t,dt)\wedge \ast
B(t,dt)=C_{\beta,N}\prod_{i<j}\sin^{2\beta}\left(\frac{t_i-t_j}{2}\right)\sum_k
\sum_{i_1<\ldots<i_k}
A_{i_1,\ldots,i_k}B_{i_1,\ldots,i_k}dt_1\wedge\cdots\wedge dt_N\,
,\end{equation}for some constant $C_{\beta,N}$.  Hence, we find that the fermionic
operators $\mc{Q}$ and $\mc{Q}^\dagger$  can be respectively interpreted
as the exterior derivative and its dual: $\mc{Q}\sim-\mathrm{i}d$ and
$\mc{Q}^\dagger\sim \mathrm{i}d^\ast$. Thus
\begin{equation}
\mc{H}=\Delta:=d\,d^\ast+d^\ast\,d\, .\end{equation} In consequence,
the Jack superpolynomials can be viewed as symmetric, homogeneous,
and orthogonal  eigenforms of a Laplace-Beltrami operator. This illustrates
the known connection between supersymmetric quantum mechanics and
 differential geometry \cite{Witten,Grandjean}. \end{remark}

  If the triangularity of the Jack  polynomial
$J_\Lambda$ with respect to the  monomial basis is imposed,
requiring that it be a common eigenfunction of  $\mathcal{H}$ and
$\mathcal{I}$ is sufficient to define it. This is our third
characterization of the Jack superpolynomials.
\begin{theorem}\cite{DLM3}  \label{TheoDefJack}
The Jack polynomials in superspace $\{J_\Lambda\}_\La$ form the
unique basis of $\mathscr{P}^{S_N}(\beta)$ such that
\begin{equation} \mc{H}(\beta)\,J_\La=\varepsilon_\La(\beta)\,
J_\La\,,\quad \mc{I}(\beta)\,J_\La=\epsilon_\La(\beta)\, J_\La\aand
J_\La=m_\La+\sum_{\Om<\La}c_{\La\Om}(\beta)m_\Om\,.\end{equation}
The eigenvalues are given explicitly by
\begin{align}
\varepsilon_\La(\beta)&=\sum_{j=1}^N[
(\Lambda_j^*)^2+\beta(N+1-2j)\Lambda_j^*]\, ,\\
\epsilon_\La(\beta)&=\sum_{i=1}^m[\Lambda_i-\beta m(m-1)-\beta
\#_\La]\, ,\end{align}
  where $\#_\La$ denotes the
number of pairs $(i,j)$ such that $\La_i<\La_j$ for $1\leq i\leq m$
and $m+1\leq j\leq N$.
\end{theorem}

When no Grassmannian variables are involved, that is when
$\overline{\underline{\La}}=0$, our characterizations of the Jack
superpolynomials specialize to known characterizations of the Jack
polynomials that can be found for instance in \cite{Stan}. However, in the usual case there is a more common characterization of the Jack polynomials
in which the scalar product appearing in Proposition~\ref{charac1}
is replaced by the  scalar product (\ref{defscalprodcomb}).
As already announced, this more combinatorial characterization can be extended to the supersymmetric case.
But before turning to the analysis of the behavior of $J_\La$ with respect to the combinatorial scalar product, we present a lemma concerning properties
of the eigenvalues $\varepsilon_\La(\beta)$ and $\epsilon_\La(\beta)$.

\begin{lemma}\label{lemmaeigenvalues1} Let $\La\vdash(n|m)$
and write $\la=\La^*$.  Let also $\varepsilon_\La(\beta)$ and $\epsilon_\La(\beta)$
be the eigenvalues given in Theorem~\ref{TheoDefJack}. Then
\begin{align} \varepsilon_\La(\beta)&=2\sum_j
j(\lambda'_j-\beta
\la_j)+\beta n(N+1)-n\, ,\\
\epsilon_\La(\beta)&=|\La^a|-\beta|{\La'}^a|-\beta\frac{m(m-1)}{2}\,
.\end{align}
  \end{lemma}
  \begin{proof}
The first formula is known (see \cite{Stan} for instance).
As
for the second one, we consider
\begin{equation}\#_\La=\sum_{i=1}^m \#_{\La_i}\, ,\end{equation}
where $\#_{\La_i}$ denotes the number of parts in $\La^s$ bigger
than $\La_i$.  But from the definition of the conjugation, we easily
find that
\begin{equation}\#_{\La_i}={\La'}_{m+1-i}+1-i\, ,\end{equation}
so that
\begin{equation}\#_{\La}=\sum_{i=1}^m
({\La'}_i+1-i)=|{\La'}^a|+\frac{m(m-1)}{2} \, ,
\end{equation}
from which the second formula follows.
\end{proof}

\section{Combinatorial orthogonality of the  Jack superpolynomials}

In terms of the   scalar product (\ref{defscalprodcomb}),  we can directly check the self-adjointness of our eigenvalue-problem defining operators, $\mc{H}$ and $\mc{I}$.

\begin{proposition}\label{propositionHI} The operators $\mc{H}$ and $\mc{I}$
defined in (\ref{defHIs}) are, when $N\to\infty$, self-adjoint (symmetric) with respect
to the  scalar product
$\LL \, \cdot \, | \, \cdot \,\RR_{\beta}$ defined in
(\ref{defscalprodcomb}).  \end{proposition}
\begin{proof}
We first rewrite the limit as $N \to \infty$
of  $\mc{H}$ and $\mc{I}$ in terms of power sums. Since
these differential operators are both of order two, it is sufficient
to determine their action on the products of the form $p_m p_n$,
$\tilde{p}_m p_n $ and $\tilde{p}_m \tilde{p}_n $.  Direct
computations give
\begin{multline}
\mc{H}=\sum_{n\geq 1}[ n^2+\beta
n(N-n)](p_n\,\partial_{p_n}+\tilde{p}_n\,\partial_{\tilde{p}_n})
+\beta \sum_{n,m\geq 1}[(m+n)p_m\,
p_n\,\partial_{p_{m+n}}+2m\,p_n\tilde{p}_m\,\partial_{\tilde{p}_{n+m}}
] \\+\sum_{n,m\geq1}m n [p_{m+n}\,\partial_{p_n}\, \partial_{p_m}+2
\tilde{p}_{n+m}\,\partial_{\tilde{p}_m} \partial_{p_n}\,]
\end{multline}
  and
\begin{multline}
    \mc{I}=\sum_{n\geq0}(1-\beta)(n \tilde{p}_n\,
\partial_{\tilde{p}_n})+\! \frac{\beta}{2}\sum_{m,n\geq 0}\tilde{p}_m\,
\tilde{p}_n\, \partial_{\tilde{p}_m}\,
\partial_{\tilde{p}_n}\\+ \sum_{m\geq 0, n\geq 1}[ n\, \tilde{p}_{m+n}\,
\partial_{\tilde{p}_m}\,
\partial_{p_n}+\beta p_n\, \tilde{p}_m\,\partial_{\tilde{p}_{m+n}}]\, .
\end{multline}
  Note that these equations are valid when $N$ is
either infinite or finite.  In the latter case, the sums over the
terms containing $\tilde{p}_m$ and $p_n$ are respectively restricted
such that $m\leq N-1$ and $n\leq N$.

Then, letting $A^\perp$ denote the adjoint of a generic operator $A$
with respect to the scalar product (\ref{defscalprodcomb}), it is easy to check that
\begin{equation}\beta\,p_n^\perp=n\partial_{p_n} \aand
\beta\,\tilde{p}_n^\perp=\partial_{\tilde{p}_n}\,.\end{equation}
Hence, comparing the three previous equations, we obtain that
$\mc{H}^\perp=\mc{H}$ and $\mc{I}^\perp=\mc{I}$. For these
calculations, we observe that $(ab)^\perp =b^\perp a^\perp$, even
when $a$ and $b$ are both fermionic.
  \end{proof}

In order to demonstrate the orthogonality
of the Jack superpolynomials with respect to the scalar product
\eqref{defscalprodcomb}, the
most natural path consists in  establishing the self-adjointness of all the
operators $\mc{H}_n$ and $\mc{I}_n$. But proceeding as for $\mc{H}$ and $\mc{I}$
above, by trying to reexpress them in terms of $p_n$, ${\tilde p}_n$ and their
derivatives, seems hopeless. An indirect line of attack is mandatory.

Let us first recall that the conserved operators (\ref{defHI}) can all be expressed in terms of
the Dunkl-Cherednik operators defined in (\ref{Dunk}). The ${\mc{D}}_i$'s commute among themselves:
\begin{equation} [{\mc{D}}_i, {\mc{D}}_j]=0\;.
\end{equation}
They obey the Hecke relations (correcting a misprint in
Eq. (25) of \cite{DLM3})
\begin{equation}\label{commu}
{\mc{D}}_i K_{i,i+1}-K_{i,i+1}\, {\mc{D}}_{i+1}  =\beta\,.\end{equation} We will also need the following commutation
relations:
\begin{equation}\label{comA}[{\mc{D}}_i,x_i] = x_i+\beta
\left(\sum_{j<i}x_iK_{ij}+\sum_{j>i}x_jK_{ij}\right)\;,\end{equation}
while if $i\not= k$,
\begin{equation}\label{comB}
[{\mc{D}}_i,x_k]= -\beta x_{{\rm max} (i,k)}\,
K_{ik}\;.\end{equation}

The idea of the proof of the orthogonality is the following: in a first step, we show that the conserved
operators ${\mc{H}}_n$ and ${\mc{I}}_n$ are self-adjoint with respect to the scalar
product \eqref{defscalprodcomb} and then  we demonstrate that this implies the orthogonality of the
$J_\La$'s.  The self-adjointness property is established via the kernel: showing that
$F= F^\perp$ is the same as showing that
\begin{equation}F^{(x)}K^{\beta,N}= F^{(y)} K^{\beta,N}\; ,
\end{equation}
where  $K^{\beta,N}$  is the restriction of $K^{\beta}$  defined in Theorem \ref{cauchyppbeta} to $N$ variables, and
where $F^{(x)}$ (resp. $F^{(y)}$) stands for the operator $F$ in the
variable $x$ (resp. $y$). In order to prove this for  our conserved
operators  ${\mc{H}}_n$ and ${\mc{I}}_n$,  we need to establish some
results on the action of symmetric monomials in the  Dunkl-Cherednik
operators  acting on the following expression:
\begin{equation}\label{defO}
{\tilde\Omega}:=\prod_{i=1}^N\frac{1}{(1-x_iy_i)}\prod_{i,j=1}^N\frac{1}{(1-x_iy_j)^\beta}\; , \end{equation}
as well as some modification of ${\tilde\Omega}$. For that
matter, we recall a result of Sahi \cite{Sahi}:
\begin{proposition}\label{sahi} The action of the Dunkl-Cherednik operators ${\mc{D}}_j$ on ${\tilde\Omega}$ defined by (\ref{defO}) satisfies:
\begin{equation}
{\mc{D}}_j^{(x)}{\tilde\Omega}= {\mc{D}}_j^{(y)}{\tilde\Omega}\;.
\end{equation}
\end{proposition}

Before turning to the core of our argument, we establish the
following lemma.
\begin{lemma}\label{lemnouv}
Given a set $J= \{j_1,\dots, j_\ell \}$, denote by $x_J$ the product
$x_{j_1} \dots x_{j_{\ell}}$.  Suppose $x_J=K_{\sigma} x_I$ for some
$\sigma \in S_N$ such that $K_{\sigma} F K_{\sigma^{-1}}=F$.  Then
\begin{equation}
\frac{1}{ x_I} F^{(x)}  x_I \, \tilde \Omega = \frac{1}{ y_I}
F^{(y)}  y_I \, \tilde \Omega \quad \implies \quad \frac{1}{ x_J}
F^{(x)}  x_J \, \tilde \Omega = \frac{1}{ y_J} F^{(y)}  y_J \,
\tilde \Omega \, .
\end{equation}
\end{lemma}
\begin{proof}  The proof is straightforward and only uses the simple property
$K_{\sigma}^{(x)} \tilde \Omega = K_{\sigma^{-1}}^{(y)} \tilde
\Omega$.  To be more precise, we have
\begin{eqnarray}
\frac{1}{x_J} F^{(x)} x_J \, \tilde \Omega &=&
K_{\sigma}^{(x)} \frac{1}{x_I} F^{(x)} x_I K_{\sigma^{-1}}^{(x)} \, \tilde \Omega  \nonumber=K_{\sigma}^{(y)} K_{\sigma}^{(x)} \frac{1}{x_I} F^{(x)} x_I  \, \tilde \Omega  \nonumber =K_{\sigma}^{(y)} K_{\sigma}^{(x)} \frac{1}{y_I} F^{(y)} y_I  \, \tilde \Omega \nonumber \\
&=&
K_{\sigma}^{(y)}  \frac{1}{y_I} F^{(y)} y_I   K_{\sigma^{-1}}^{(y)} \, \tilde \Omega  = \frac{1}{y_J} F^{(y)} y_J \, \tilde \Omega \, .
\end{eqnarray}
\end{proof}

We are now ready to attack the main proposition.

\begin{proposition}\label{autoco}
The mutually commuting  operators ${\mc{H}}_n$ and ${\mc{I}}_n$
satisfy
\begin{equation}\label{autocha}{\mc{H}}_n^{(x)}K^{\beta,N}= {\mc{H}}_n^{(y)} K^{\beta,N}
\qquad {\rm and}\qquad  {\mc{I}}_n^{(x,\theta)}K^{\beta,N}= {\mc{I}}_n^{(y,\phi)}
K^{\beta,N}\;,\end{equation}
with $K^{\beta,N}$ the restriction to $N$ variables of the
kernel $K^\beta$ defined in Theorem \ref{cauchyppbeta}.
\end{proposition}

\begin{proof}  We first expand the kernel as follows:
\begin{eqnarray}
 K^{\beta,N} &=& K_0 \prod_{i,j}\left(1+\beta \frac{\theta_i \phi_j}{(1-x_i y_j)}\right) \\
&=&
 K_0\left\{ 1+\beta e_1 \left(
\frac{\ta_i\phi_j}{(1-x_iy_j)} \right) + \cdots + \beta^N e_N
\left(\frac{\ta_i\phi_j}{(1-x_iy_j)}\right)
 \right\}
\end{eqnarray}
where $K_0$ stands for $K^{\beta,N}(x,y,0,0)$, {\it i.e.},
\begin{equation}
 K_0:=\prod_{i,j=1}^N\frac{1}{(1-x_iy_j)^\beta}\; ,
\end{equation}
and where $ e_{\ell} (u_{i,j}) $ is the elementary symmetric
function $e_\ell$ in the variables
\begin{equation}
u_{i,j}:= \frac{\ta_i\phi_j} {(1-x_iy_j)} \qquad  i,j=1,\dots,N\; .
\end{equation}
Note that, in these variables,
the maximal possible elementary symmetric function is $e_N$ given that
$\theta_i^2=\phi_i^2=0$. In the following, we will use the compact
notation $I^-=\{1,\cdots, i-1\}$ and $I^+=\{i,\cdots, N\}$ (and
similarly for $J^\pm$), together with  $w_{I^-}= w_1\cdots w_{i-1}$
and $w_{I^+}= w_i\cdots w_N$.

The action of the operators on $K^\beta$ can thus be decomposed into
their action on each monomial in this expansion.  Now observe that
$K_0$ is invariant under the exchange of any two variables $x$ or
any two variables $y$. Therefore, if an operator $F$ is such that
${\mathcal K}_{\sigma} F {\mathcal K}_{\sigma}^{-1}=F$ for all
$\sigma \in S_N$, and such that
\begin{equation} \label{eqaprouver}
F^{(x,\theta)} v_{I^-} \, K_0 = F^{(y,\phi)} v_{I^-} \, K_0\; \qquad
{\rm with}\qquad v_i:=u_{i,i}\end{equation} for all $i=1,\dots,N+1$,
then we immediately have by symmetry that $F^{(x,\theta)}
K^{\beta} =  F^{(y,\phi)} K^{\beta}$.  We will use this
observation in the case of $\mathcal H_n$ and $\mathcal I_n$.

We first consider the case $F=\mathcal H_n$. Recall from (\ref{defHI}) that
${\mc{H}}_n =p_n({\mc{D}}_i) $
is such that ${\mathcal K}_{\sigma} {\mathcal H}_n {\mathcal
K}_{\sigma}^{-1}={\mathcal H}_n$ (see \cite{DLM3}).  Since $\mathcal
H_n$ does not depend on the fermionic variables, we thus have to
prove from the previous observation that
\begin{equation}
{\mathcal H}_n^{(x)}  \frac{1}{(1-xy)_{I^-} }\, K_0 = {\mathcal
H}_n^{(y)} \frac{1}{(1-xy)_{I^-} } \, K_0\;, \end{equation} or
equivalently
\begin{equation}
{\mathcal H}_n^{(x)}  \,(1-xy)_{I^+}\, \,{\tilde\Omega} = {\mathcal
H}_n^{(y)} \,(1-xy)_{I^+}\, {\tilde\Omega}\;,
\end{equation}
for all $i=1,\dots,N+1$ (the case $i=N+1$ corresponds to the empty
product).

The underlying symmetry of the problem implies the result will follow from
showing
\begin{equation} y_{J^+} \, {\mathcal H}_n^{(x)} \, x_{J^+}  {\tilde\Omega} =
 x_{J^+} \, {\mathcal H}_n^{(y)} \, y_{J^+}  \,{\tilde\Omega}\;,
\end{equation} for $j\geq i$, or equivalently,
\begin{equation}\frac{1}{x_{J^+}  } {\mathcal H}_n^{(x)}\, {x_{J^+}
  }\,{\tilde\Omega} =
\frac{1}{y_{J^+}  } {\mathcal H}_n^{(y)}\, {y_{J^+}
}\,{\tilde\Omega}\;.
\end{equation}
This follows from Lemma~\ref{lemnouv} which assures us
 that all the different terms can be obtained
from these special ones.

Now, instead of analyzing the family ${\mathcal
H}_n=p_n({\mc{D}}_i)$, it will prove simpler to  consider the
equivalent  family $e_n({\mc{D}}_i)$.  We will first show the case
$e_N({\mc{D}}_i)$; that is,
\begin{equation}\frac{1}{x_{J^+} } {\mc{D}}^{(x)}_1\cdots {\mc{D}}^{(x)}_N\, x_{J^+}
\, {\tilde\Omega} =\frac{1}{y_{J^+} }{\mc{D}}^{(y)}_1\cdots
{\mc{D}}^{(y)}_N \, y_{J^+} \, {\tilde\Omega}\;.\end{equation} Let
us concentrate on the left hand side. We note that
\begin{equation}
 \frac{1}{x_{J^+} } {\mc{D}}^{(x)}_1\cdots {\mc{D}}^{(x)}_N \, x_{J^+} \,
{\tilde{\Omega}} = \frac{1}{x_{J^+} } {\mc{D}}^{(x)}_1{x_{J^+}
}\cdots\frac{1}{x_{J^+} } {\mc{D}}^{(x)}_N\, x_{J^+}
{\tilde{\Omega}}\;. \end{equation} It thus suffices to
study each term $({x_{J^+} })^{-1} {\mc{D}}_j{x_{J^+} }$ separately.
In each case we find that
\begin{equation} {\mc{D}}_k\, {x_{J^+} } = {x_{J^+} } \, {\tilde {\mc{D}}}_k\;.\end{equation} The
form of ${\tilde {\mc{D}}}$ depends upon $j$ and $k$. There are two
cases:
\begin{eqnarray}\label{castr}
 k<j: &&{\tilde {\mc{D}}}_k= {\mc{D}}_k - \beta\sum_{\ell=j}^N K_{\ell,k}\;,\cr
k\geq j:&& {\tilde {\mc{D}}}_k= {\mc{D}}_k+ 1+
\beta\sum_{\ell=1}^{j-1} K_{\ell,k}\;
\end{eqnarray}
which can be easily checked using (\ref{comA}) and (\ref{comB}).
We can thus write
\begin{equation}\frac{1}{x_{J^+} } {\mc{D}}^{(x)}_1\cdots {\mc{D}}^{(x)}_N
\, x_{J^+}  {\tilde\Omega}
 = {\tilde {\mc{D}}}^{(x)}_1\cdots {\tilde {\mc{D}}}^{(x)}_N\,{\tilde\Omega}\;.\end{equation}
Using Proposition \ref{sahi} and $K_{ij}^{(x)} \tilde \Omega =K_{ij}^{(y)} \tilde \Omega$, the rightmost term ${\tilde
{\mc{D}}}^{(x)}_N$ can thus be changed into ${\tilde {\mc{D}}}^{(y)}_N$. Since it commutes with the previous terms (i.e.,
it acts on the variables $y$ while the others act on $x$), we have
\begin{eqnarray}
 {\tilde {\mc{D}}}^{(x)}_1\cdots {\tilde {\mc{D}}}^{(x)}_{N-1}{\tilde
{\mc{D}}}^{(y)}_N \,{\tilde\Omega} & =&{\tilde {\mc{D}}}^{(y)}_N
{\tilde {\mc{D}}}^{(x)}_1\cdots {\tilde {\mc{D}}}^{(x)}_{N-1}
\,{\tilde\Omega}=  {\tilde {\mc{D}}}^{(y)}_N {\tilde
{\mc{D}}}^{(y)}_{N-1}\cdots {\tilde {\mc{D}}}^{(y)}_{1}
\,{\tilde\Omega}\cr &=& \frac{1}{y_{J^+} }{\mathcal D}^{(y)}_N
y_{J^+} \cdots \frac{1}{y_{J^+} }{\mathcal D}^{(y)}_1 \, y_{J^+}
{\tilde\Omega}\cr\ &=& \frac{1}{y_{J^+} }{\mathcal D}^{(y)}_N\cdots
{\mathcal D}^{(y)}_1 \, y_{J^+} {\tilde\Omega} =\frac{1}{y_{J^+} }{\mathcal D}^{(y)}_1\cdots {\mathcal D}^{(y)}_N \,
y_{J^+} {\tilde\Omega}\;,\end{eqnarray} which is the desired result.

At this point, we have only considered a single conserved operator,
namely $e_N({\mathcal D}_i)$. But by replacing ${\mc{D}}_i$ with
${\mc{D}}_i+t$ in $e_N({\mathcal D}_i)$, we obtain a generating
function  for all the operators $e_n({\mathcal D}_i)$.  Since to
prove $e_N({\mathcal D}_i^{(x)}+t) K^{\beta,N}= e_N({\mathcal
D}_i^{(y)}+t) K^{\beta,N}$ simply amounts to replacing ${\tilde
{\mc{D}}}_i$ by ${\tilde {{\mc{D}}}}_i+t$ in the previous argument,
we have completed the proof of ${\mc{H}}_n^{(x)}K^{\beta,N}={\mc{H}}_n^{(y)} K^{\beta,N}$.

For the case of ${\mc{I}}_n$, we start with the expression given in (\ref{defHI})
which readily implies that  ${\mathcal K}_\sigma \,
{\mathcal I}_n \, {\mathcal K}_\sigma^{-1}= {\mathcal I}_n$.
Therefore, from the observation surrounding formula
(\ref{eqaprouver}), and because the derivative $\ta_1 \d_{\ta_1}$
annihilates the $K_0$ term in the expansion of $K^{\beta,N}$, we only
need to show that
\begin{equation}
{\mathcal I}_n^{(x,\ta)} v_{I^-}\, K_0 = {\mathcal I}_n^{(y,\phi)}
 v_{I^-} \,  K_0\;, \end{equation}
for $i=2,\dots,N+1$. Up to an overall multiplicative factor, the
only contributing part in ${\mc{I}}_n$, when acting on $ v_{I^-} $, is
\begin{equation}
{\mc O}_n:= {\mc{D}}_1^n+K_{12}{\mc{D}}_1^nK_{12} + \cdots
K_{1,i-1}{\mc{D}}_1^nK_{1,i-1}\;.\end{equation} It thus suffices to show
that
\begin{eqnarray}
{\mc O}_n^{(x)}\, (1-xy)_{I^+}\, {\tilde\Omega}
 =  {\mc O}_n^{(y)}\, (1-xy)_{I^+} \, {\tilde\Omega}
\end{eqnarray}
Once more, we can use Lemma~\ref{lemnouv} since ${\mc O}_n$ commutes
with $K_{k,\ell}$ for $k,\ell\geq i$. Thus, we only need to check that
for $j\geq i$,
\begin{eqnarray}
\frac{1}{x_{J^+} } {\mc O}_n^{(x)}{x_{J^+} }\,{\tilde\Omega}
=\frac{1}{y_{J^+} } {\mc O}_n^{(y)}{y_{J^+} }\,{\tilde\Omega}\;.
\end{eqnarray}
Since the $K_{1\ell}$'s act trivially on the  variables $x_j$ for
$j> \ell$, the previous relation reduces to proving
\begin{equation}
\frac{1}{x_{J^+} } [{\mc{D}}_1^n]^{(x)} \, {x_{J^+}
}\,{\tilde\Omega} = \frac{1}{y_{J^+} }[{\mc{D}}_1^n]^{(y)}\,
{y_{J^+}}\,{\tilde\Omega}\;.\end{equation} The left hand side takes
the form
\begin{equation}
\frac{1}{x_{J^+} } [{\mc{D}}_1^n]^{(x)} \, {x_{J^+}
}\,{\tilde\Omega}= \left\{ \frac{1}{x_{J^+} }{\mc{D}}_1^{(x)}\,
{x_{J^+} }\right\}^n \,{\tilde\Omega}\;.\end{equation} We then only
have to evaluate $({x_{J^+}})^{-1}{\mc{D}}_1^{(x)}\, {x_{J^+}}$. The
result is given by the first case in (\ref{castr}) (since $j>1$) .
The proof is completed as follows
\begin{eqnarray}
\left\{ \frac{1}{x_{J^+}}{\mc{D}}_1^{(x)}\, {x_{J^+}}\right\}^n
\,{\tilde\Omega}=  [{\tilde {\mc{D}}}^{(x)}_1]^n \,{\tilde\Omega}=   [{\tilde {\mc{D}}}^{(y)}_1]^n \,{\tilde\Omega} = \frac{1}{y_{J^+} }
[{\mc{D}}_1^n]^{(y)}\,  {y_{J^+}
}\,{\tilde\Omega}\;.\end{eqnarray}\end{proof}

As previously mentioned, the proposition has the following
corollary.
\begin{corollary}\label{autocos}
The operators $\mc{H}_r$ and $\mc{I}_s$
defined in (\ref{defHI}) are self-adjoint (symmetric) with respect
to the  scalar product
$\LL \, \cdot \, | \, \cdot \, \RR_{\beta,N}$
given in (\ref{defscalprodcombN}).  \end{corollary}
This immediately gives our main result.
\begin{theorem}\label{mainortho} The Jack superpolynomials $\{J_{\La} \}_{\La}$ are orthogonal
with respect to the combinatorial scalar product; that is,
\begin{equation}
\LL J_{\La} | J_{\Om} \RR_{\beta} \propto \delta_{\La,\Om}\;.
\end{equation}
\end{theorem}

\begin{proof} The fact that in $N$  variables
$\LL J_{\La} | J_{\Om} \RR_{\beta,N} \propto \delta_{\La,\Om}$
is a consequence of Corollary~\ref{autocos} and
Proposition~\ref{charac2}, which says that the Jack superpolynomials are the
unique common eigenfunctions of the $2N$ operators appearing in
Corollary~\ref{autocos}.  Given that the
expansion coefficients of the Jack superpolynomials in terms of
supermonomials do not depend on the number of variables $N$ \cite{DLM3},
the theorem then follows from (\ref{degN}).
\end{proof}

\begin{remark}
  That
the Jack superpolynomials are orthogonal with respect to the
analytical and combinatorial
scalar products is certainly remarkable given their rather different nature.
Even in the absence of fermionic variables, the orthogonality of the
Jack polynomials with respect to both scalar products is a highly
non-trivial observation. In that case, one can provide a partial
rationale for the compatibility between the two scalar products, by
noticing their equivalence in the following two circumstances
\cite{Kadell, Mac}:
\begin{equation}\label{equivbeta1}
 \langle f  | g \rangle_{\beta=1,N}=\LL f  | g \RR_{\beta=1,N}\qquad (m=0)
\end{equation}
(see e.g., \cite{Mac} VI.9 remark 2) and
\begin{equation}\label{equivNinfty}\lim_{N \rightarrow\infty}\frac{\langle f | g
\rangle_{\beta,N}}{\langle 1|1\rangle_{\beta,N}}= \LL f | g
\RR_\beta \, ,\qquad (m=0) \end{equation} (see e.g., \cite{Mac} VI.9 (9.9)) for
$f,\, g,$ two arbitrary symmetric polynomials.
In superspace, when $m\not=0$, this compatibility between the two products is even
more remarkable since the limiting-case equivalences
\eqref{equivbeta1} and \eqref{equivNinfty} are simply lost. This is
most easily seen by realizing that, after integration over the
fermionic variables, we obtain
 \begin{equation} \langle p_\la
\tilde{p}_n | p_\mu \tilde{p}_r \rangle_{\beta,æN}=\langle p_\la  |
p_\mu p_{r-n}\rangle_{\beta,æN}\, ,\quad r>n\, ,\end{equation} and
thus the power sums cannot be orthogonal for any value of $N$ and
$\beta$. This shows that the connection between the two scalar
products is rather intricate.\end{remark}

\begin{corollary} \label{routes} The following statements are direct consequences of the orthogonality
property of the Jack polynomials in superspace.
\begin{enumerate}
\item  The Jack polynomials in superspace $\{J_{\La}\}_{\La}$
form the unique basis of
$\mathscr{P}^{S_\infty}$
such that \begin{equation}\begin{array}{lll} 1)& J_{\Lambda}= m_{\Lambda} +\sum_{\Om < \La}
c_{\La \Om }(\beta) m_{\La}&\mbox{(triangularity)};\\
2)&\LL  J_{\La}| J_{\Om} \RR_\beta \propto
\delta_{\La,\Om}&\mbox{(orthogonality)}.\end{array}\end{equation}
\item Let $K^\beta$ be the reproducing kernel defined in
Theorem~\ref{cauchyppbeta}.  Then,
\begin{equation}K^\beta(x,\theta;y,\phi)=\sum_{\La\in\mathrm{SPar}}j_{\La}(\beta)^{-1}\versg{J_\Lambda(x,\theta)}\,
\versd{J_\Lambda(y,\phi)}\,.\end{equation} where
\begin{equation}\label{normde}
j_\La(\beta):=\LL
\versg{J_{\La}} | \versd{J_{\La}} \RR_{\beta}\;.
\end{equation}
\item Let $\{g_{\Lambda}\}_{\Lambda}$ be the basis, defined in (\ref{defgla}),
dual to that of the monomials
with respect to the combinatorial scalar product.
Then, the Jack superpolynomials expand upper  triangularly in
this basis:
\begin{equation}
J_{\La} =\sum_{\Om \geq \La} u_{\La \Om}(\beta) \, g_{\Om} \, ,
\quad {\rm with} \quad u_{\La \La}(\beta) \neq 0 \, .
\end{equation}
\end{enumerate}
\end{corollary}
\begin{proof}
{\it 1}.  We have seen that the Jack polynomials in superspace satisfy 1) and 2).
To prove unicity, suppose $\{\tilde J_{\Lambda} \}_\La$ satisfies 1) and 2).
It was shown in \cite{DLM3} that the operators $\mathcal
H$ and $\mathcal I$ act triangularly on the  monomial basis.  Thus,
$\mathcal H$ and $\mathcal I$ also act triangularly on the basis $\{
\tilde J_{\Lambda} \}_\La$. Furthermore, from
Proposition~\ref{propositionHI}, they are self-adjoint with respect
to the combinatorial scalar product.  Hence, we must conclude from
the orthogonality of $\{\tilde J_{\Lambda} \}_\La$
that  $\tilde J_{\La}$ is an eigenfunction of
$\mathcal H$ and $\mathcal I$, from which Theorem~\ref{TheoDefJack}
implies that $\tilde J_{\La} =J_{\La}$.

\bigskip

\n {\it 2}. The proof is similar to that of  Lemma~\ref{cauchybasesbeta} (see also Section VI.2 of
\cite{Mac}).

\bigskip

\n {\it 3}.  Suppose that $\LL J_{\La}| J_{\Om} \RR_\beta \propto
\delta_{\La,\Om}$, and let $J_{\La}= \sum_{\Om \in \mathcal S}
u_{\La \Om} \, g_{\Om}$, where $\mathcal S$ is some undefined set.
If $\La$ is not the smallest element of $\mathcal S$, then there
exists at least one element $\Gamma$ of $\mathcal S$ that does not
dominate any other of its elements. In this case, we have
\begin{equation}
\LL J_{\La}| J_{\Gamma} \RR_\beta= \sum_{\Om \in \mathcal S} u_{\La
\Om}(\beta) \sum_{\Delta \leq \Gamma} c_{\Gamma \Delta}(\beta) \,
\LL g_{\Om}| m_{\Delta} \RR_\beta \, .
\end{equation}
Since $\Gamma$ does not dominate any element of $\mathcal S$, the
unique non-zero contribution in this expression is that of
  $u_{\La \Gamma}(\beta) \, c_{\Gamma \Gamma}(\beta)
\,  \LL g_{\Gamma}, m_{\Gamma} \RR_\beta= u_{\La \Gamma}(\beta)$.
Since this term is non-zero by supposition, we have the
contradiction $0=\LL J_{\La}| J_{\Gamma} \RR_\beta=u_{\La
\Gamma}(\beta)\neq 0$.
\end{proof}

Actually, it
can be shown that all statements of Corollary~\ref{routes} and
Theorem~\ref{theoduality} below are not only consequences of
Proposition~\ref{mainortho} but are equivalent to it.

\section{Further properties}

\subsection{Duality}

In this subsection, we show that the homomorphism
$\hat{\omega}_\beta$, defined in Eq.~(\ref{defendobeta}), has a
simple action on Jack superpolynomials.   To avoid any confusion, we
make explicit the $\beta$ dependence   by writing
$J^{(1/\beta)}_\La$.

\begin{remark} The rationale for this notation is to match the one used in \cite{Mac} when $m=0$: $J^{(1/\beta)}_{\La}(x,\ta)= J^{(1/\beta)}_{\La^s}(x)= J^{(\alpha)}_{\La^s}(x)$, where
$\alpha=1/\beta$. (Similarly, in our previous works
\cite{DLM1,DLM3}, we denoted $J^{(1/\beta)}_\La$ by
$J_\La(x,\theta;1/\beta)$ to
keep our definition similar to the usual form introduced by Stanley
\cite{Stan} as $J_\la(x;\alpha)$ when $m=0$). We stress however, that when we need to make explicit the $\beta$-dependence of $j_\La,\, \mc{H}$ and $\mc{I}$, we write $j_\La(\beta)\,,\mc{H}(\beta)$and $\mc{I}(\beta)$ respectively.\end{remark}

\begin{proposition}\label{propomega}
One has
\begin{equation}\mc{H}(\beta)\hat{\omega}_{\beta}J_\La^{(\beta)}=\varepsilon_{\La'}(\beta)\,
\hat{\omega}_{\beta}J_\La^{(\beta)}\aand
\mc{I}(\beta)\hat{\omega}_{\beta}J_\La^{(\beta)}=\epsilon_{\La'}(\beta)\,
\hat{\omega}_{\beta}J_\La^{(\beta)}\, .\end{equation}
\end{proposition}
\begin{proof}Let us rewrite the special form of the operator
$\mc{H}(\beta)$ appearing in the proof of Proposition~\ref{propositionHI} as
\begin{equation}\mc{H}(\beta)=\sum_{n\geq1}[n^2+\beta
n(N-n)]\hat{A}_n+\sum_{m,n\geq 1}(\beta
\hat{B}_{m,n}+\hat{C}_{m,n})\, ,\end{equation} with
\begin{equation}\begin{array}{lll}
  \hat{A}_n&=&p_n\,
\partial_{p_n}+\tilde{p}_n\,
\partial_{\tilde{p}_n}\, ,\\
\hat{B}_{m,n}&=&(m+n)p_m\, p_n\,
\partial_{p_{m+n}}+2m\,p_n\tilde{p}_m\,
\partial_{\tilde{p}_{n+m}}\, ,\\
\hat{C}_{m,n}&=&m n  \left( p_{m+n}\,
\partial_{p_n}\, \partial_{p_m}+2 \tilde{p}_{n+m}\,\partial_{\tilde{p}_m}
\partial_{p_n} \right)\, .
\end{array}\end{equation}From these definitions, we get
\begin{equation}\hat{\omega}_{1/\beta}\hat{A}_n=\hat{A}_n \,
\hat{\omega}_{1/\beta}\, ,
\quad
\hat{\omega}_{1/\beta}\hat{B}_{m,n}=-\frac{1}{\beta}\hat{B}_{m,n} \,
\hat{\omega}_{1/\beta} \aand
\hat{\omega}_{1/\beta}\hat{C}_{m,n}=-{\beta}\hat{C}_{m,n} \,
\hat{\omega}_{1/\beta} \, .\end{equation} These relations imply
\begin{eqnarray}
\hat{\omega}_{1/\beta}\mc{H}(\beta)\hat{\omega}_{\beta}=\sum_{n\geq1}[n^2+\beta
n(N-n)]\hat{A}_n-\! \! \! \! \sum_{m,n\geq 1}(
\hat{B}_{m,n}+\beta\hat{C}_{m,n})
=(1+\beta)N\sum_{n\geq1}n\hat{A}_n-\beta\mc{H}(1/\beta)\, . \nonumber
\end{eqnarray}Now,
  considering  $\sum_{n\geq1}n\hat{A}_nm_\La=|\La|m_\La$ and Lemma
\ref{lemmaeigenvalues1}, we obtain
\begin{equation}\hat{\omega}_{1/\beta}\mc{H}(\beta)
\hat{\omega}_{\beta}J_\La^{(\beta)}=\varepsilon_{\La'}(\beta)J_\La^{(\beta)}\,
\end{equation} as claimed. The relation involving $\mc{I}(\beta)$ is
proved in a similar way.
\end{proof}

For the next theorem, we will need the following result from \cite{DLM6}:

\begin{proposition} \label{theoebase}Let $\La$ be a superpartition and $\La'$ its conjugate. Then \begin{equation}\label{einm}
\versg{ e_\La}=m_{\La'}+\sum_{\Om
< \La'} N_{\La}^{\Om}\,m_\Om\, ,\qquad {\rm with } \qquad N_{\La}^{\Om}\in \Z\;. \end{equation}
\end{proposition}

\begin{theorem}\label{theoduality} The homomorphism $\hat{\omega}_{\beta}$ is such that
\begin{equation}\hat{\omega}_{1/\beta}\versd{J^{(1/\beta)}_\La}=j_\La(\beta)\versg{J^{(\beta)}_{\La'}}\,\;,
\end{equation}
with $ j_\La(\beta)$ defined in (\ref{normde}).
\end{theorem}
\begin{proof}
Let us first prove that $\hat{\omega}_{\beta}J_\La^{(\beta)}\propto J_{\La'}^{(1/\beta)}$.
 From the third point of Corollary~\ref{routes},  we know that $J_{\La}^{(1/\beta)}
=\sum_{\Om \geq \La} u_{\La \Om}(\beta) \, g_{\Om}$. But
Eq.~(\ref{dualityge}) implies $\hat{\omega}_{1/\beta}g_\La=e_\La$.
Hence,
\begin{equation}
\hat \omega_{1/{\beta}} \Bigl(J_\La^{(1/\beta)}\Bigr) = \sum_{\Om \geq
\La} u_{\La \Om}(\beta) \, e_{\Om} = \sum_{\Om \geq \La} u_{\La
\Om}(\beta) \sum_{\Gamma \leq \Om'} N_{\Om}^{\Gamma} \,  \versg{m_{\Gamma}}=  \sum_{\Gamma \leq \La'} v_{\La \Gamma}(\beta) \, \versg{m_{\Gamma} }\, ,
\end{equation}
where we have used (\ref{dualityge}), Proposition~\ref{theoebase} and
the fact that $\Om \geq \La \iff \Om' \leq \La'$. Further, since
$N_{\Lambda}^{\Lambda'}=1$ and $u_{\La \La}(\beta)\neq 0$, we have
$v_{\La \La'} \neq 0$. Now, from Proposition~\ref{propomega},
$\hat{\omega}_{{1/\beta}} \Bigl(J_\La^{(1/\beta)}\Bigr)$ is an
eigenfunction of $\mathcal H(1/\beta)$ and $\mathcal I(1/\beta)$
with eigenvalues $\varepsilon_{\La'}(1/\beta)$ and
$\epsilon_{\La'}(1/\beta)$ respectively.   The triangularity we just
obtained ensures from Theorem~\ref{TheoDefJack}, that
$\hat{\omega}_{{1/\beta}} \Bigl(J_\La^{(1/\beta)}\Bigr)$ is
proportional to ${J_{\La'}^{(\beta)}}$.

Again from Proposition~\ref{theoebase}, we know that
$m_\La=(-1)^{m(m-1)/2}e_{\La'}+\mbox{higher terms}$, so that
\begin{equation}
J^{(1/\beta)}_\La=(-1)^{m(m-1)/2}e_{\La'}+\mbox{higher terms}\,
.\end{equation} Moreover,  from Eq.~\eqref{dualityge}, we get
\begin{equation} \hat{\omega}_\beta J_\La^{(1/\beta)}=(-1)^{m(m-1)/2}g_{\La'}+\mbox{higher terms}\, .
\end{equation}
But  the proportionality proved above  implies
\begin{equation}
\hat{\omega}_{1/\beta}\versd{J^{(1/\beta)}_\La} =A_\La(\beta)\, \versg{J_{\La'}^{(\beta)} }=A_\La(\beta)\, \versg{m_{\La'}}Ì°å°Ì­åÉ+\mbox{lower
terms}\, ,
\end{equation} for some constant $A_\La(\beta)$.  Finally,
considering the duality between $g_\La$ and $m_\La$, we obtain
\begin{eqnarray}
(-1)^{m(m-1)/2}j_\La(\beta)&=&\LL  J_\La^{(1/\beta)} |
J_\La^{(1/\beta)}\RR_\beta \nonumber\\
&=& \LL \hat{\omega}_{\beta}
J_\La^{(1/\beta)} | \hat{\omega}_{1/\beta} J_\La^{(1/\beta)}\RR_\beta \nonumber\\
&=& \LL (-1)^{m(m-1)/2} \versd{g_{\La'} }| A_\La(\beta)\versg{m_{\La'}}\RR_\beta \nonumber\\
&=& (-1)^{m(m-1)/2} A_\La(\beta)
\end{eqnarray} as desired.
\end{proof}

\subsection{Limiting cases}

In Section 5, we have proved that the Jack superpolynomials
are orthogonal with respect to the combinatorial scalar product.
This provides a direct link with the classical symmetric functions
in superspace. Other links,  less general but more explicit, are
presented in this section, from the consideration of $J_\La$ for
special values of $\beta$ or for particular superpartitions.

\begin{proposition}  \label{propogjack} For $\La= (n)$ or $(n;0)$, one has
(using the notation of Proposition~\ref{propgn}):
 \begin{equation}J_{(n)}=\frac{n!}{(\beta+n-1)_n}g_n
\aand J_{(n;0)}=\frac{n!}{(\beta+n)_{n+1}}\tilde{g}_n\,
.\end{equation}
\end{proposition}
\begin{proof}
Since $J_{(0;1^n)}=m_{(0;1^n)}=\tilde{e}_n$, we have on the one hand
$\hat{\omega}_{\beta}(J_{(0;1^n)}) =\tilde g_n$ from
(\ref{dualityge}). On the other hand, from
Proposition~\ref{propomega}, $\hat{\omega}_{\beta}(J_{(0;1^n)})$ is
an eigenfunction of $\mathcal H(\beta)$ and $\mathcal I(\beta)$ with
eigenvalues $\varepsilon_{(n;0)}(\beta)$ and
$\epsilon_{(n;0)}(\beta)$ respectively. Since $(n;0)$ is the highest
partition with one fermion in the order on superpartitions, we have from Theorem
\ref{TheoDefJack}, that there exists a unique eigenfunction of
$\mathcal H$ and $\mathcal I$ with such eigenvalues. We must thus
conclude that $\tilde g_n$ is also proportional to
$J_{(n;0)}$. Looking at Proposition \ref{propgn}
and considering that the coefficient of $m_{(n;0)}$ in $J_{(n;0)}$
needs to be equal to one, we obtain
$(\beta+n)_{n+1}J_{(n;0)}={n!}\,\tilde{g}_n$.  The relation between
$J_{(n)}$ and $g_n$ is proved in a similar way.
\end{proof}

\begin{corollary}\label{coronorme} For $\La= (n)$ or $(n;0)$, the combinatorial norm of $J_\La$ is
\begin{equation} \LL J_{(n)}|J_{(n)}\RR_\beta=\frac{n!}{(\beta+n-1)_n}\aand
\LL J_{(n;0)}|J_{(n;0)}\RR_\beta=\frac{n!}{(\beta+n)_{n+1}}\,
.\end{equation}
\end{corollary}
\begin{proof}Using the previous proposition, we get
\begin{align}
(n!)^2\,\LL
g_n|g_n\RR_\beta=&(\beta+n-1)_n^{2}\;\LL J_{(n)}|J_{(n)}\RR_\beta\,
,\cr
 (n!)^2\,\LL
\tilde{g}_n|\tilde{g}_n\RR_\beta=&(\beta+n)_{n+1}^{2}\;\LL
J_{(n;0)}|J_{(n;0)}\RR_\beta\, .
 \end{align}
From
Proposition~\ref{propgn}, we know that
\begin{equation}n!\,g_n=(\beta+n-1)_n \; m_{(n)}+\ldots \, ,\quad
n!\,\tilde{g}_n=(\beta+n)_{n+1}\; m_{(n;0)}+\ldots \, ,\end{equation}
where the dots stand for lower terms in the order on superpartitions.  Thus,
considering Corollary \ref{corodualitygm}, we get
\begin{equation} \LL g_n|g_n\RR_\beta=\frac{(\beta+n-1)_n}{n!}\,
,\quad \LL
\tilde{g}_n|\tilde{g}_n\RR_\beta=\frac{(\beta+n)_{n+1}}{n!}\end{equation}
and the proof follows.
\end{proof}

\begin{theorem}\label{limdebeta}
For $\beta=0,\,1,$ or $\beta\rightarrow \infty$, the limiting expressions of $J_\La^{(1/\beta)}$ are
\begin{equation}J_\La^{(1/\beta)}\longrightarrow\left\{\begin{array}{ll}
m_\La\, & {\rm when ~}\beta\longrightarrow0\, ,\\
\versg{e_{\La'} } \, &{\rm when ~} \beta\longrightarrow\infty\,
.\end{array}\right.\end{equation}
and
\begin{equation}J_{(n)}^{(1)}=h_n\aand
J_{(n;0)}^{(1)}=\frac{1}{n+1}\tilde{h}_n\, .\end{equation}
\end{theorem}
\begin{proof}The case $\beta \to 0$
is a direct consequence of Theorem \ref{TheoDefJack}, given that
$\mathcal H(\beta)$ and $\mathcal I(\beta)$ act diagonally on
supermonomials in this limit. The second case is also obtained from
the eigenvalue problem. Indeed, when $\beta\rightarrow\infty$,
  $\beta^{-1}\mc{H}(\beta)$ and  $\beta^{-1}\mc{I}(\beta)$ behave as
first order differential operators.
  Then, it is easy to get
  \begin{equation}\bigg[\lim_{\beta\rightarrow\infty}\frac{\mc{H}(\beta)}{\beta}\bigg]\,
  e_{\La'}=\bigg[-2\sum_j j\la_j+n(N-1)\bigg]\,e_{\La'}\wwhere
\la=\La^*\end{equation}
($\La^*$ being defined in Lemma~\ref{lemmaeigenvalues1}) and
  \begin{equation}\bigg[\lim_{\beta\rightarrow\infty}\frac{\mc{I}(\beta)}{\beta}\bigg]\,
  e_{\La'}=\bigg[-|\La^a|-\frac{m(m-1)}{2}\bigg]\,e_{\La'}\, .\end{equation}
These are the eigenvalues of $J_{\La}$ in the limit where $\beta\rightarrow \infty$ (cf. Lemma
\ref{lemmaeigenvalues1}). The proportionality constant between  $e_{\La'}$ and $J_\La$ is fixed by   Proposition \ref{theoebase} and Theorem \ref{TheoDefJack}.  We have thus
 \begin{equation} \versg{e_{\La'} }  =
\lim_{\beta\rightarrow\infty} \versd{J_\La^{(1/\beta)} }\, .\end{equation}
  Finally, we note that the property concerning $h_n$ and
$\tilde{h}_n$ is an immediate
  corollary of Proposition~\ref{propogjack}.
  \end{proof}

\subsection{Normalization}

In this subsection,  $\tilde m_{\Lambda}$ shall denote the augmented supermonomial:
\begin{equation}
\tilde m_{\Lambda} = n_{\Lambda}! \, m_{\Lambda} \, ,
\end{equation}
where $n_{\Lambda}!$ is defined in Proposition~\ref{propgn}.

It is easy to see that the smallest
superpartition of degree $(n|m)$ in the order on superpartitions is
\begin{equation}\label{minide}
\Lambda_{\mathrm{min}}:=(\delta_m\,;\,
1^{\ell_{n,m}}\,) \, ,\end{equation} where
\begin{equation}\label{delnm}
\delta_m:=(m-1,m-2,\ldots,0)\, ,  \quad \ell_{n,m}:=n-|\delta_m| \quad \aand
|\delta_m|=\frac{m(m-1)}{2}\, .\end{equation} Now, let
$c_\La^{\mathrm{min}}(\beta)$ stand for the coefficient of
$\tilde m_{\Lambda_{\mathrm{min}}}$ in the monomial expansion of
$J_\La^{(1/\beta)}$. We will establish a relation between this
coefficient and the norm of the Jack superpolynomials $J_\La$.

\begin{proposition}The norm $j_\La(\beta)$  defined  in (\ref{normde}), with $\La\vdash(n|m)$, is
\begin{equation}
j_\La(\beta)= \beta^{-m-\ell_{n,m}}
 \, \frac{c_\La^{\mathrm{min}}(\beta)}{c_{\La'}^{\mathrm{min}}(1/\beta)} \end{equation}
\end{proposition}
\begin{proof}
One readily shows that
\begin{equation}
m_{\La_\mathrm{min}}=p_{\La_\mathrm{min}}+\mbox{higher
terms}\,.\end{equation} Since $m_{\La_\mathrm{min}}$ is the only
supermonomial containing $p_{\La_\mathrm{min}}$, we can write
\begin{equation}\label{pmini}
J_{\La}^{(1/\beta)}=c_\La^{\mathrm{min}}(\beta) \, p_{\La_\mathrm{min}}+\mbox{higher
terms}\, .\end{equation}
Let us now apply $\hat{\omega}_{1/\beta}$ on this expression. Using  Eq.~\eqref{involugen} we get
\begin{equation} \hat{\omega}_{1/\beta}{J^{(1/\beta)}_\La}=\beta^{-m-\ell_{n,m}}(-1)^{m(m-1)/2}\, {c_\La^{\mathrm{min}}(\beta)}\,
p_{\La_\mathrm{min}}+\mbox{higher terms}\, .\end{equation}
But if we apply $\hat{\omega}_{1/\beta}$ on $J_\La^{(1/\beta)}$ by using first
Theorem~\ref{theoduality} to write it as $(-1)^{m(m-1)/2 }\,  j_\La(\beta)J_{\La'}^{(\beta) }$ and expand $J_{\La'}^{(\beta)}$ using (\ref{pmini}), we get instead
\begin{equation}\hat{\omega}_{1/\beta}{J^{(1/\beta)}_\La}=j_\La(\beta)(-1)^{m(m-1)/2}\, {c_{\La'}^{\mathrm{min}}(1/\beta)}\, p_{\La_\mathrm{min}}+\mbox{higher
terms}\;.
\end{equation}
Here we have used the fact that $\Lambda_{\mathrm{min}}$, being the smallest
superpartition of degree $(n|m)$ in the ordering on superpartitions,
labels the smallest supermonomial in both the decomposition of $J_\La$ and $J_{\La'}$. The result follows from the comparison of the last two equations.
\end{proof}

The coefficient $c_\La^{\mathrm{min}}(\beta)$ appears from computer
experimentation to have a very simple form. We will  now introduce
the notation needed to describe it. Recall (from the definition of
conjugation in Section~2) that $D[\Lambda]$ is the diagram  used to
represent $\Lambda$. Given a cell $s$ in $D[\Lambda]$, let
$a_{\La}(s)$ be the number of cells (including the possible circle
at the end of the row) to the right of $s$.  Let also
$\ell_{\La}(s)$ be the number of cells (not including the possible
circle at the bottom of the column) below $s$.  Finally, let
$\Lambda^{\circ}$, be the set of cells of $D[\La]$ that do not
appear at the same time in a row containing a circle {\it and} in a
column containing a circle.
\begin{conjecture}  The coefficient  $c_\La^{\mathrm{min}}(\beta)$ of
$\tilde m_{\Lambda_{\mathrm{min}}}$
in the monomial expansion of $J_\La^{(1/\beta)}$ is given by
\begin{equation}
c_\La^{\mathrm{min}}(\beta) = \frac{1}{\prod_{s \in
\Lambda^{\circ}} \Bigl( a_{\Lambda}(s)/\beta + \ell_{\Lambda}(s)+1
\Bigr)  }
\end{equation}
with $ {\La_\mathrm{min}}$ is defined in (\ref{minide}).
\end{conjecture}
For instance, if $\Lambda=(3,1,0;4,2,1)$, we can fill $D[\La]$ with
the values $\bigl( a_{\Lambda}(s)/\beta + \ell_{\Lambda}(s)+1
\bigr)$ corresponding to the cells $s \in \La^{\circ}$.  This gives
(using $\gamma=1/\beta$):
\begin{equation} {{\tableau[mcY]{{\mbox{\tiny
$3\gamma+5$}}&{\mbox{\tiny $2\gamma+3$}} &{\mbox{\tiny $\gamma+2$}}&
{\mbox{\tiny $1$}} \\& &{\mbox{\tiny
$\gamma+1$}}&\bl\gcercle\\{\mbox{\tiny $\gamma+3$}}&{\mbox{\tiny
$1$}}\\ & \bl\gcercle\\{\mbox{\tiny $1$}}\\  \bl \gcercle}} }
\end{equation}
Therefore, in this case,
\begin{equation}
c_\La^{\mathrm{min}}(\beta) =\frac{1}{(3/\beta+5)(2/\beta+3)(1/\beta+2)(1/\beta+1)(1/\beta+3)}
\end{equation}

Even though the Jack superpolynomials cannot be normalized to have
positive coefficients when expanded in terms of monomials, we
nevertheless conjecture they satisfy the following {\it integrality}
property.
\begin{conjecture} Let
\begin{equation}
J_\Lambda^{(1/\beta)}= c_\La^{\mathrm{min}}(\beta)\sum_{\Omega \leq
\Lambda} \tilde{c}_{\Lambda \Omega}(\beta) \, \tilde m_{\Omega}.
\end{equation}
Then $\tilde{c}_{\Lambda \Omega}$ is a polynomial in $1/\beta$ with
integral coefficients.
\end{conjecture}

\section{Outlook: Macdonald polynomials in superspace}

In this work, we have highlighted the existence of a
one-parameter (i.e., $\beta$) deformation
of the scalar product as the key tool for defining Jack
superpolynomials combinatorially. However,
there  also exists a two-parameter deformation ($t$ and $q$) of the
combinatorial scalar product. Again, this has a natural lift to the
superspace, namely
\begin{equation}  \label{qtspro}\LL \versg{p_\La}|
\versd{p_\Om}\RR_{q,t}:=z_\La(q,t)\delta_{\La,\Omega}\,
,\end{equation}
where
\begin{equation}
z_\La(q,t)=z_\La \prod_{i=1}^{m}\frac{1-q^{\La_i+1}}{1-t^{\La_i+1}}
\prod_{i=m+1}^{\ell(\La)}\frac{1-q^{\La_i}}{1-t^{\La_i}}\, ,\quad
m=\overline{\underline{\La}}\, .\end{equation}
This reduces to the previous scalar product $ \LL
\cdot| \cdot\RR_\beta$ when $q=t^{1/\beta}$ and $t\rightarrow 1$.
The generalized form of the reproducing kernel reads
\begin{equation}
\prod_{i,j}\frac{\left(tx_iy_j+t\theta_i\phi_j;q\right)_\infty}{\left(x_iy_j+\theta_i\phi_j;q\right)_\infty}
=\sum_\La
z_\La(q,t)^{-1}\versg{p_\La(x,\theta)}\versd{p_\La(y,\phi)}\,
,\end{equation}
with $(a;q)_\infty:=\prod_{n\geq 0}(1-aq^n)$.

Now, the scalar product (\ref{qtspro}) leads directly to a
conjectured  definition of Macdonald
superpolynomials.

\begin{conjecture}
In the  space of symmetric superfunctions with rational coefficients
in
  $q$ and $t$, there exists a basis $\{  M_\La\}_{\La}$, where
  $M_\La=M_\La(x,\theta;q,t)$,
such that \begin{equation}(1)\quad M_{\Lambda} =m_{\Lambda}
+\sum_{\Om < \La}
  C_{\La \Om }(q,t) m_{\La}\qquad\mbox{and}\qquad (2)
\quad\LL  \versg{M_{\La}}|\versd{ M_{\Om}} \RR_{q,t} \propto
\delta_{\La,\Om}\,.\end{equation}
\end{conjecture}

Note that in this context, the combinatorial construction cannot be
compared with the analytical one since the corresponding
supersymmetric eigenvalue problem has not been formulated yet. In
other words, the  proper supersymmetric version of the
Ruijsenaars-Schneider model \cite{Ruij} is still missing.

\begin{acknow}
 This work was  supported by NSERC and
FONDECYT (Fondo Nacional de Desarrollo Cient\'{\i}fico y
Tecnol\'ogico) grant \#1030114. P.D. is grateful to NSERC for a
postdoctoral  fellowship.
\end{acknow}

\end{document}